\documentclass[]{interact}

\usepackage{epstopdf}
\usepackage[caption=false]{subfig}

\usepackage[numbers,sort&compress]{natbib}
\graphicspath{{figuras/}}
\bibpunct[, ]{[}{]}{,}{n}{,}{,}

\usepackage{algorithm}  
\usepackage{algpseudocode}
\usepackage{url} 
\usepackage{adjustbox}
\usepackage{multirow}
\usepackage{hyperref}
\hypersetup{
	colorlinks=true,
	linkcolor=blue,
	citecolor=blue,
	urlcolor=blue
}

\renewcommand\bibfont{\fontsize{10}{12}\selectfont}
\theoremstyle{plain}

\theoremstyle{definition}

\theoremstyle{remark}

\begin{document}

\title{A Joint Bayesian Boolean Matrix Factorization with Application to Chromosomal Copy Number Alterations in Multiple Myeloma}

\author{
\name{Adolphus Wagala\textsuperscript{a,b}\thanks{CONTACT A. Wagala. Email: adolphus@ds.dfci.harvard.edu} and Samur Mehmet\textsuperscript{a,b} and Giovanni Parmigiani\textsuperscript{a,b}}
\affil{\textsuperscript{a}Department of Data Science, Dana Farber Cancer Institute, 450 Brookline Ave., Boston, 02115, MA, USA; \textsuperscript{b} {Department of Biostatistics, Harvard University, 677 Huntington Ave., 02115, Boston, MA,USA}}
}

\maketitle

\begin{abstract}
Boolean matrix factorization provides an interpretable framework for discovering latent binary patterns in high-dimensional data, yet existing methods typically analyze a single binary matrix or factorize multiple matrices independently, failing to exploit shared latent structure across related datasets. We propose Joint Bayesian Boolean Matrix Factorization (\texttt{JBBMF}), a model that simultaneously factorizes two related binary matrices through a shared latent Boolean pattern matrix and dataset-specific loading matrices. To capture dependence between paired datasets, we introduce a conditional prior linking the loading matrices, allowing latent factors to persist or change across conditions while preserving a common interpretable representation. The model combines Boolean matrix factorization with a Bernoulli observation model and conjugate priors, yielding closed-form full conditional distributions and an efficient Gibbs sampler for posterior inference, uncertainty quantification for latent factors, reconstructed matrices, and noise parameters. Simulation studies demonstrate that jointly modeling related binary datasets substantially improves recovery of shared latent factors compared with independently applying standard Boolean matrix factorization to each dataset, while maintaining high reconstruction accuracy. We apply \texttt{JBBMF} to paired chromosomal copy number alteration profiles from multiple myeloma patients collected at diagnosis and relapse. The analysis identifies recurrent chromosomal alteration signatures shared between disease stages and quantifies the uncertainty of these findings. \texttt{JBBMF} offers a flexible and interpretable Bayesian model for the joint analysis of related binary datasets in genomics and other application domains.
\end{abstract}


\section{Introduction}
\label{sec:Introduction}
Binary matrices arise naturally in a wide range of contemporary  applications. In computational biology, copy-number profiles and  mutation catalogs are routinely encoded as binary indicators of  genomic events across patients; in information retrieval, user--item interactions are recorded as binary presence-or-absence matrices; and in network analysis, each relation type in a multi-relational network yields a binary adjacency matrix. Analyzing such data requires methods that respect their discrete combinatorial structure. Boolean matrix factorization (\texttt{BooMF}) provides a natural framework for this purpose: it approximates a binary matrix as a Boolean product of two low-rank binary matrices, one encoding latent patterns over features and the other recording which patterns are active for each observation. Because all quantities remain binary throughout, \texttt{BooMF} operates on the original scale of the data and yields factors with a direct, interpretable meaning that real-valued alternatives such as principal component analysis (PCA) or non-negative matrix factorization (NMF) do not provide. PCA produces dense continuous loadings that are difficult to interpret when the binary encoding carries intrinsic meaning; NMF allows arbitrary positive real values that may be equally uninformative in a binary context. Section~\ref{sec:Basic_Concepts} treats \texttt{BooMF} formally and reviews existing algorithms.

However, in many contemporary applications, data do not arrive as a single matrix. Instead, they can be organized into multiple related binary matrices: paired datasets collected at different time points (such as chromosomal copy-number profiles at diagnosis and at relapse in multiple myeloma), multi-relational networks in which each relation type defines a separate binary matrix, or heterogeneous data sources that share common entities such as patients, genes, or experimental conditions. Applying \texttt{BooMF} independently to each matrix in such settings ignores the shared structure that connects them, risks learning incomparable latent factors, and discards weak but consistent patterns that only emerge when all matrices are analyzed together. Uncovering latent factors shared across related binary matrices therefore calls for a joint Boolean matrix factorization approach.

Relatively little work addresses this problem directly. The most closely related contribution is that of \cite{miettinen2012finding}, who extended the classical \texttt{BooMF} to a Joint Subspace Boolean Matrix Factorization (\texttt{JSBMF}) model. \texttt{JSBMF} assumes that each matrix admits a Boolean factorization in which part of the latent structure lies in a common subspace---capturing shared patterns---while the remaining components explain view-specific variation, and extracts these components via a greedy search algorithm. Despite its conceptual appeal, the \texttt{JSBMF} has important limitations. Inference is entirely deterministic and optimization-based, so the method yields only point estimates with no uncertainty quantification and no principled basis for model comparison beyond reconstruction error. The greedy strategy, while computationally feasible, offers no guarantees of global optimality and can be sensitive to initialization and factor selection order. Crucially, the formulation assumes exact or near-exact Boolean structure without an explicit noise model, reducing robustness in high-noise or sparse regimes and making the framework difficult to extend to hierarchical or multi-view settings. A second line of work, due to \cite{krmelova2013boolean}, extends Boolean factor analysis to multi-relational data by first factorizing each matrix independently using formal concept-based algorithms such as \texttt{GreConD} \cite{Belohlavek2010jcss}, then linking the resulting factors across matrices via compatibility criteria. While interpretable multi-relational Boolean structures can be recovered in this way, the linkage is post-hoc: it does not rely on a shared latent representation and does not involve a unified inference scheme that jointly learns factors across all matrices.

No existing method combines joint Boolean matrix factorization with a fully Bayesian treatment. A Bayesian formulation is particularly valuable here for three reasons. First, it provides a noise model that accommodates the imperfect Boolean structure typically present in real data, including measurement error and biological variability. Second, it yields posterior uncertainty over both the shared patterns and their dataset-specific usages, enabling the analyst to distinguish stable findings from those driven by noise. Third, it admits the incorporation of prior biological or domain knowledge, for example, about the expected frequency of chromosomal gains at a given disease stage, through the prior distribution. We therefore propose a Joint Bayesian Boolean Matrix Factorization (\texttt{JBBMF}) model in which two related binary matrices are generated from a shared latent Boolean factor matrix and dataset-specific loading matrices. The shared factor matrix captures recurring patterns common to both datasets, while the loading matrices capture how those patterns are expressed in each dataset separately. The model embeds this structure within a fully probabilistic generative framework: each observed entry follows a Bernoulli distribution whose success probability depends on the corresponding Boolean product entry, and conjugate priors on all unknown quantities yield closed-form full conditional distributions. Posterior inference is therefore carried out via an efficient Gibbs sampler that is straightforward to implement and scales to moderately large binary datasets.

The remainder of this paper is organized as follows. Section~\ref{sec:Basic_Concepts} introduces basic concepts including Boolean matrix factorization, reviews existing algorithms, and provides a brief background of multiple myeloma and copy number alterations as a motivation for a joint Boolean matrix factorization approach. Section~\ref{sec:jBBMF} presents the \texttt{JBBMF} model, the prior specification, and the Gibbs sampler. The posterior uncertainty quantification is discussed in Section~\ref{sec:uncertainty}. Simulation experiments are presented in Section~\ref{sec:sim_Expts}, while the corresponding results are presented in Section~\ref{sec:sim_results}. Section~\ref{jBBMF_real_data} presents the application to real CNA data. The summary and conclusions are given in Section~\ref{summ_conclusions}.

\section{Basic Concepts}
\label{sec:Basic_Concepts}
\subsection{Boolean Matrix Factorization}
\label{subsec:BooMF}

Let $\mathbf{X} \in \{0,1\}^{K \times G}$ be a binary matrix, where $K$ is the number of observations and $G$ is the number of features. Boolean matrix factorization (\texttt{BooMF}) seeks binary matrices $\mathbf{W} \in \{0,1\}^{K \times R}$ and $\mathbf{H} \in \{0,1\}^{R \times G}$, with $R \leq \min(K,G)$, such that $\mathbf{X}$ can be approximated by the Boolean product $\mathbf{W} \circ \mathbf{H}$, defined entrywise as
\begin{equation}
	Z_{kg}= \bigvee_{r=1}^R \big( W_{kr} \wedge H_{rg} \big),
\end{equation}
where $\wedge$ denotes logical conjunction \texttt{(AND)}, $\vee$ denotes logical disjunction \texttt{(OR)}, and the indices range over $k = 1,\dots,K$ and $g = 1,\dots,G$. In other words, $Z_{kg}= 1$ if there exists at least one latent factor $r$ such that $W_{kr} = 1$ and $H_{rg} = 1$; otherwise $Z_{kg}= 0$.

When $\mathbf{Z} = \mathbf{X}$, we say that $\mathbf{X}$ admits an exact Boolean factorization of rank $R$. In practice, however, $\mathbf{X}$  may contain noise or deviations from any low-rank Boolean structure, so we aim for an approximate factorization. A common approach is to minimize a Boolean reconstruction error such as the Hamming distance or the number of mismatched entries between $\mathbf{X}$ and $\mathbf{W} \circ \mathbf{H}$. In our CNA context, each row of $\mathbf{H}$ corresponds to recurrent binary patterns (or signatures) of activation of entries in $\mathbf{X}$. The entries of $\mathbf{H}$ indicate which features are included in each pattern. The rows of $\mathbf{W}$ describe how these patterns are used to explain each observation: entry $W_{kr} = 1$ indicates that pattern $r$ is active in observation $k$. The Boolean product then asserts that a feature $g$ is active in observation $k$ if it appears in at least one of the patterns that are active in that observation. This representation is intrinsically discrete and often promotes sparsity. If most entries of $\mathbf{W}$ and $\mathbf{H}$ are zero, then each observation is explained by a small number of patterns, and each pattern involves a limited subset of features. 

\subsection{Existing Algorithms for Boolean Matrix Factorization}
Several deterministic and probabilistic algorithmic approaches have been proposed for Boolean matrix factorization (\texttt{BooMF}). Among the deterministic algorithms,  \texttt{Asso} \cite{Asso_4479462} greedily selects candidate factors from frequent association patterns, \texttt{GreConD/GreConD+} \cite{Belohlavek2010jcss} find the smallest rank below a reconstruction threshold and correspond to formal concepts; and an alternating-minimization approach by \cite{zhang2007binary} updates  $\mathbf{W}$ and $\mathbf{H}$ in turn under surrogate objective functions that simplify the exact \texttt{BooMF} loss. Although these methods differ in how they construct the factorization, they share an important limitation: they provide point estimates and do not directly account for uncertainty in the recovered factors. Probabilistic alternatives  include  \texttt{OrMachine} (a sigmoid-link generative model with Metropolis-within-Gibbs inference)  introduced  by \cite{Rukat2017pmlr}, a family mean-parameterized Boolean factor models of \cite{lumbreras2020bayesian} equipped with Dirichlet and Beta priors that impose structural sparsity constraints on the factor matrices, and the single-matrix Bayesian Boolean Matrix Factorization  (\texttt{BBMF}) by  \cite{wagala2026bbmf}  that places conjugate priors on all unknown quantities, yielding closed-form full conditional distributions and allowing posterior inference via an efficient Gibbs sampler. Although these probabilistic models can quantify uncertainty, like their deterministic counterparts, they are designed to factorize one matrix at a time. Consequently, when two related matrices are analyzed, the shared structure between them is not directly incorporated into the factorization. The proposed \texttt{JBBMF} extends the single-matrix \texttt{BBMF} model of \cite{wagala2026bbmf} to jointly factorize two related matrices using a shared latent Boolean factor matrix while retaining matrix-specific loading matrices and full conjugacy. 

\subsection{Motivation: Multiple Myeloma and Copy Number Alterations}
\label{subsec:copy_number_alterations}
\begin{figure}[tbp]
	\centerline{\includegraphics[width=.9\textwidth]{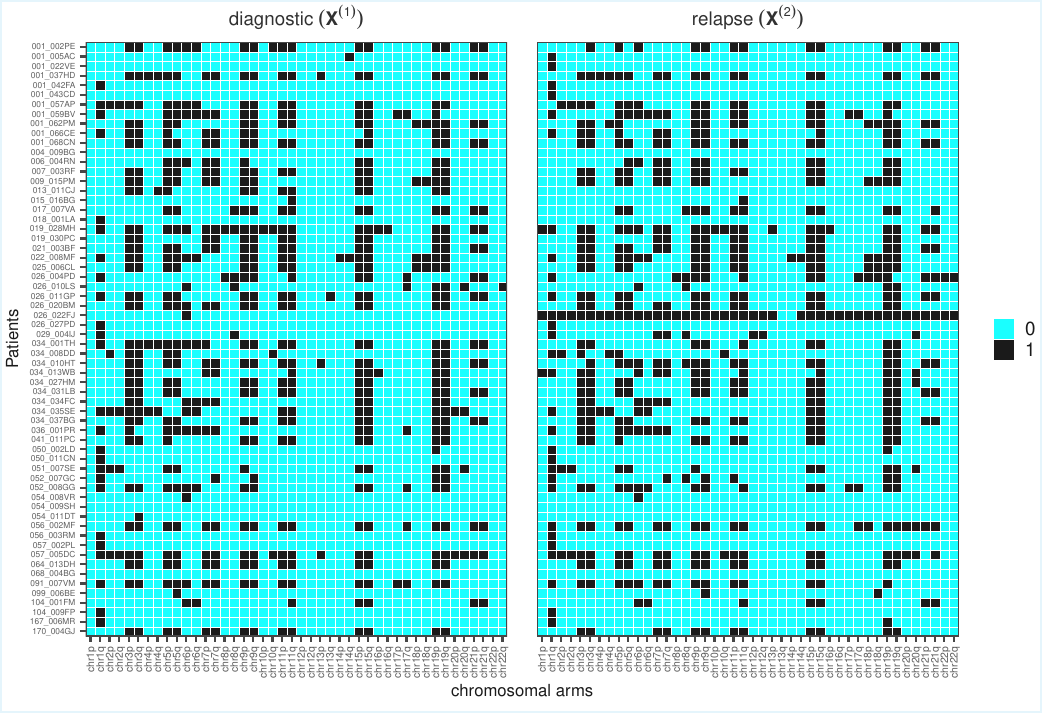}}
	\caption{Binary CNA profiles (amplification vs.\ non-amplification) at diagnosis and relapse in multiple myeloma patients. Heatmaps show copy-number states for $K=62$ patients (rows) across $G=44$ chromosomal arms (columns). Vertical bands indicate recurrently altered arms, horizontal patterns reflect overall CNA burden in individual patients, and block-like structures suggest potential patient subgroupings. Comparison between the two stages reveals both shared CNA patterns and stage-specific differences, motivating the use of a joint factorization model.}
	\label{fig:cna_diag_relapse}
\end{figure}

As a motivating application, we focus on copy number alterations (CNAs) in multiple myeloma, a malignancy of plasma cells that usually arises in the bone marrow, characterized by the clonal proliferation of abnormal plasma cells leading to bone destruction, anemia, renal dysfunction, and other clinical manifestations, and remains largely incurable for many patients despite therapeutic advances \cite{clarke2024chromosomal, mikulasova2022chromosomal, ramasamy2021100808}. Recurrent chromosomal gains and deletions are central to its pathogenesis and are associated with distinct molecular subtypes and prognostic risk \cite{cardona2021genetic, lannes2023multiple}. The CNA data analyzed here were reported by \cite{samur2023high} (paired diagnosis–relapse samples, HDM treatment context) and discretized following \cite{shen2016facets}: an arm is coded as a gain $(1)$ if its average copy number exceeds $2.2$ and a deletion $(-1)$ if below $1.8$, with the range $[1.8, 2.2]$ coded normal $(0)$; the ternary result is then binarized into amplification $(1)$ versus non-amplification $(0)$. The data comprise $K = 62$ patients profiled at diagnosis $\mathbf{X}^{(1)} \in \{0,1\}^{K \times G}$  and relapse $\mathbf{X}^{(2)} \in \{0,1\}^{K \times G}$  with $G = 44$ chromosomal arms, and  $X^{(i)}_kg = 1$, denoting amplification of arm $g$ in patient $k$ at stage $i$. Figure~\ref{fig:cna_diag_relapse} shows the observed data: vertical bands of recurrently altered arms, block-like patient subgroups, and horizontal variation in overall CNA burden are all visible. It is worth noting that  some alterations persist between stages while others appear or intensify at relapse, reflecting clonal evolution.

The vertical bands and block structures in Figure~\ref{fig:cna_diag_relapse}  suggest that a small number of shared latent signatures underlie both disease stages. Fitting each stage in isolation risks learning incomparable signatures and discarding weak-but-consistent cross-stage patterns; naively stacking the two matrices ignores the pairing and conflates stage effects with inter-patient heterogeneity. A joint factorization avoids both problems: a shared pattern matrix $\mathbf{H}$ enforces common signatures across stages, while stage-specific loading matrices $\mathbf{W}^{(1)}, \mathbf{W}^{(2)}$ control how those signatures are expressed at diagnosis and relapse. The Bayesian formulation additionally supports sparsity priors and posterior uncertainty quantification for both the signatures and their patient-level usage (Section~\ref{sec:jBBMF}).

\section{A Joint Bayesian Boolean Matrix Factorization}
\label{sec:jBBMF}
\subsection{Proposed Model}
\label{subsec:prop_model}
The proposed model, \texttt{(JBBMF)} assumes that multiple related binary matrices share a common pattern matrix $\mathbf{H}$, while each dataset retains a dataset-specific 
usage matrix $\mathbf{W}^{(i)}$, $i \in \{1, \ldots, M\}$. The $i$-th data matrix $\mathbf{X}^{(i)}$ is approximated as
\begin{align*}
	\mathbf{X}^{(i)} \approx \mathbf{Z}^{(i)} = 
	\mathbf{W}^{(i)} \circ \mathbf{H} 
	\in \{0,1\}^{K \times G},
\end{align*}
where $\circ$ denotes Boolean matrix multiplication. The entry of $\mathbf{Z}^{(i)}$ at cell $(k,g)$ is
\begin{align*}
	Z_{kg}^{(i)} = \bigvee_{r=1}^R 
	\left( W_{kr}^{(i)} \wedge H_{rg} \right),
\end{align*}
where $R \in \mathbb{Z}_+$ is the factorization rank, $\wedge$ denotes logical conjunction \texttt{(AND)} and $\vee$ denotes logical disjunction \texttt{(OR)}, so that $Z_{kg}^{(i)} = 1$ if and only if there exists at least one factor $r$ for which both $W_{kr}^{(i)} = 1$ and $H_{rg} = 1$.

Since $Z_{kg}^{(i)}$ is a deterministic function of $(\mathbf{W}^{(i)}, \mathbf{H})$, all randomness in $X_{kg}^{(i)}$ arises from a Bernoulli noise model: conditionally on $Z_{kg}^{(i)}$, each observation $X_{kg}^{(i)}$ follows a Bernoulli distribution with success probability $p_{i,11}$ if $Z_{kg}^{(i)} = 1$ and 
$p_{i,10}$ if $Z_{kg}^{(i)} = 0$. We write the generative model as
\begin{align}
	p\!\left(X^{(i)}_{kg} \mid Z^{(i)}_{kg}\right) =
	\begin{cases}
		p_{i,11}^{\,X^{(i)}_{kg}}\,(1-p_{i,11})^{\,1-X^{(i)}_{kg}}, 
		& Z^{(i)}_{kg} = 1,\\[4pt]
		p_{i,10}^{\,X^{(i)}_{kg}}\,(1-p_{i,10})^{\,1-X^{(i)}_{kg}}, 
		& Z^{(i)}_{kg} = 0,
	\end{cases}
	\label{Generative_Model2}
\end{align}
for $i \in \{1, \ldots, M\}$. The noise parameters satisfy the identifiability constraint
\begin{align}
	p_{i,11} > p_{i,10}, \quad i = 1, \ldots, M,
	\label{eq:identifiability}
\end{align}
which ensures that $Z_{kg}^{(i)} = 1$ corresponds to the signal state rather than the noise state. This constraint is not explicitly enforced in the sampler. Instead, it is treated as an assumption that holds naturally when the data contain a genuine Boolean signal. To see why, note that when the factorization is well specified, cells where $Z_{kg}^{(i)} = 1$ tend to coincide with $X_{kg}^{(i)} = 1$ more frequently than cells where $Z_{kg}^{(i)} = 0$ do, so that the sufficient statistics satisfy $n_1/n > m_1/m$ in expectation. The Beta full conditionals for $p_{i,11}$ and $p_{i,10}$ then concentrate in the region satisfying~\eqref{eq:identifiability} without additional intervention. In practice, we recommend verifying that $\hat{p}_{i,11} > \hat{p}_{i,10}$ holds throughout post-burn-in iterations as a routine diagnostic check; this was confirmed for all chains in all experiments reported in this paper.

The model factors element-wise, giving the matrix-specific likelihood components:
\begin{align*}
	p\!\left(\mathbf{X}^{(i)} \mid \mathbf{W}^{(i)}, \mathbf{H}\right)
	= \prod_{k=1}^K \prod_{g=1}^G 
	p\!\left(X^{(i)}_{kg} \mid Z^{(i)}_{kg}\right), 
	\qquad i \in \{1, \ldots, M\}.
\end{align*}
At each cell $(k,g)$, observations from different datasets are conditionally independent given the shared latent structure. 
	The full likelihood for the two-matrix case is
	\begin{align*}
		\begin{split}
			&p\!\left(\mathbf{X}^{(1)}, \mathbf{X}^{(2)} \mid 
			\mathbf{W}^{(1)}, \mathbf{W}^{(2)}, \mathbf{H}\right)
			=\\
			& \prod_{k=1}^K \prod_{g=1}^G \Bigg\{
			\left[p_{1,11}^{\,X_{kg}^{(1)}} 
			(1-p_{1,11})^{1-X_{kg}^{(1)}}\right]^{Z_{kg}^{(1)}}
			\left[p_{1,10}^{\,X_{kg}^{(1)}} 
			(1-p_{1,10})^{1-X_{kg}^{(1)}}\right]^{1-Z_{kg}^{(1)}}\\
			&\quad\times
			\left[p_{2,11}^{\,X_{kg}^{(2)}} 
			(1-p_{2,11})^{1-X_{kg}^{(2)}}\right]^{Z_{kg}^{(2)}}
			\left[p_{2,10}^{\,X_{kg}^{(2)}} 
			(1-p_{2,10})^{1-X_{kg}^{(2)}}\right]^{1-Z_{kg}^{(2)}}
			\Bigg\}.
		\end{split}
	\end{align*}
	
	\subsection{Prior Specifications}
	\label{subsec:prior}
	For the diagnostic matrix $\bm{X}^{(1)}$ and relapse matrix $\bm{X}^{(2)}$, we introduce dependence between the two datasets at the latent factor level via a conditional prior $p(\bm{W}^{(2)} \mid \bm{W}^{(1)})$, so that each $W^{(2)}_{kr}$ depends on the corresponding $W^{(1)}_{kr}$ at each row--factor pair $(k,r)$. For each  $(k,r)$ and $w \in \{0,1\}$ entries are bernoulli:
	\begin{align*}
		p\!\big(W^{(2)}_{kr}=w \mid W^{(1)}_{kr}=1\big) 
		&= \gamma_{11}^{\,w}(1-\gamma_{11})^{\,1-w},\\
		p\!\big(W^{(2)}_{kr}=w \mid W^{(1)}_{kr}=0\big) 
		&= (1-\gamma_{00})^{\,w}\,\gamma_{00}^{\,1-w}.
	\end{align*}
	Sufficient statistics are counts, defined using the notation  $\{W^{(1)}_{kr}=a,\,W^{(2)}_{kr}=b\}$ with $a,b \in \{0,1\}$, as follows:
	\begin{align*}
		&n_{11}=\#\{(k,r):W^{(1)}_{kr}=1,\,W^{(2)}_{kr}=1\},\quad
		n_{10}=\#\{(k,r):W^{(1)}_{kr}=1,\,W^{(2)}_{kr}=0\},\\
		&n_{01}=\#\{(k,r):W^{(1)}_{kr}=0,\,W^{(2)}_{kr}=1\},\quad
		n_{00}=\#\{(k,r):W^{(1)}_{kr}=0,\,W^{(2)}_{kr}=0\}.
	\end{align*}
	so that, as is customary, 
	\begin{align}
		p(\bm{W}^{(2)} \mid \bm{W}^{(1)}, \gamma_{11}, \gamma_{00})
		= \gamma_{11}^{\,n_{11}}(1-\gamma_{11})^{\,n_{10}}\,
		\gamma_{00}^{\,n_{00}}(1-\gamma_{00})^{\,n_{01}}.
	\end{align}
	Table~\ref{tab:priors} collects the complete prior specification, including the
	observation-noise parameters introduced in Section~\ref{subsec:prop_model}.
	\begin{table}[tbp]
		\centering
		\small
		\caption{Prior specification for \texttt{JBBMF}. Each latent binary quantity carries a
			Bernoulli prior whose rate carries a Beta hyper-prior. The
			final column gives the corresponding full conditional derived in
			Appendix~\ref{app:conditionals}.}
		\label{tab:priors}
		\begin{tabular}{llp{0.26\textwidth}lc}
			\toprule
			Quantity & Prior & Role & Hyperparameters & Full cond. \\
			\midrule
			\multicolumn{5}{l}{\emph{Diagnosis loadings}}\\
			\midrule
			$W^{(1)}_{kr}\mid\alpha^{(1)}_k$\label{prior:W1} & $\mathrm{Bernoulli}(\alpha^{(1)}_k)$
			& Factor $r$ active for patient $k$ at diagnosis
			& --- & \eqref{cond:W1} \\
			$\alpha^{(1)}_k$\label{prior:alpha} & $\mathrm{Beta}(a_1,a_2)$
			& Patient-specific activation rate
			& $a_1,a_2$ & \eqref{cond:alpha} \\
			\addlinespace
			\multicolumn{5}{l}{\emph{Relapse loadings (transition from diagnosis)}}\\
			\midrule
			$W^{(2)}_{kr}\mid W^{(1)}_{kr}=1$ & $\mathrm{Bernoulli}(\gamma_{11})$
			& Factor active at diagnosis persists to relapse
			& --- & \eqref{cond:W2} \\
			$W^{(2)}_{kr}\mid W^{(1)}_{kr}=0$ & $\mathrm{Bernoulli}(1-\gamma_{00})$
			& Factor inactive at diagnosis is newly acquired
			& --- & \eqref{cond:W2} \\
			$\gamma_{11}$ & $\mathrm{Beta}(u_{11},v_{11})$
			& Persistence probability
			& $u_{11},v_{11}$ & \eqref{cond:gamma11} \\
			$\gamma_{00}$ & $\mathrm{Beta}(u_{00},v_{00})$
			& Probability of remaining inactive
			& $u_{00},v_{00}$ & \eqref{cond:gamma00} \\
			\addlinespace
			\multicolumn{5}{l}{\emph{Shared pattern matrix}}\\
			\midrule
			$H_{rg}\mid\beta_g$\label{prior:H} & $\mathrm{Bernoulli}(\beta_g)$
			& Chromosomal arm $g$ belongs to factor $r$
			& --- & \eqref{cond:H} \\
			$\beta_g\mid\psi_g=1$\label{prior:beta} & $\mathrm{Beta}(b_1,b_2)$
			& Slab: arm $g$ broadly relevant across factors
			& $b_1,b_2$ & \eqref{cond:beta} \\
			$\beta_g\mid\psi_g=0$ & $\mathrm{Beta}(c_1,c_2)$
			& Spike: arm $g$ shrunk towards exclusion; requires
			$b_1/b_2>c_1/c_2$
			& $c_1,c_2$ & \eqref{cond:beta} \\
			$\psi_g$\label{prior:psi} & $\mathrm{Bernoulli}(\pi)$
			& Indicator that arm $g$ is broadly relevant
			& --- & \eqref{cond:psi} \\
			$\pi$\label{prior:pi} & $\mathrm{Beta}(d_1,d_2)$
			& Mixing weight, so sparsity of $\mathbf{H}$ is learned
			& $d_1,d_2$ & \eqref{cond:pi} \\
			\addlinespace
			\multicolumn{5}{l}{\emph{Observation noise, diagnosis ($i=1$)}}\\
			\midrule
			$p_{1,11}$ & $\mathrm{Beta}(1,1)$
			& $\Pr(X^{(1)}_{kg}=1\mid Z^{(1)}_{kg}=1)$
			& $1,1$ & \eqref{distP1_11} \\
			$p_{1,10}$ & $\mathrm{Beta}(1,1)$
			& $\Pr(X^{(1)}_{kg}=1\mid Z^{(1)}_{kg}=0)$
			& $1,1$ & \eqref{cond:p110} \\
			\midrule
			\multicolumn{5}{l}{\emph{Observation noise, relapse ($i=2$)}}\\
			\midrule
			$p_{2,11}$ & $\mathrm{Beta}(1,1)$
			& $\Pr(X^{(2)}_{kg}=1\mid Z^{(2)}_{kg}=1)$
			& $1,1$ & \eqref{distP2_11} \\
			$p_{2,10}$ & $\mathrm{Beta}(1,1)$
			& $\Pr(X^{(2)}_{kg}=1\mid Z^{(2)}_{kg}=0)$
			& $1,1$ & \eqref{cond:p210} \\
			\bottomrule
		\end{tabular}
	\end{table}
	
	\subsection{Posterior Distribution}
	\label{subsec:post_dist}
	Having specified the likelihood and prior components of the model with the conditional prior on $\mathbf{W}^{(2)}$, the joint posterior distribution is obtained by combining all unknown quantities 
	\begin{align*}    
		\big(\mathbf{W}^{(1)},\, \mathbf{W}^{(2)},\,\mathbf{H},\, \boldsymbol{\alpha}^{(1)},\, 
		p_{1,11},\, p_{1,10},\, p_{2,11},\, p_{2,10},\,\gamma_{11},\, \gamma_{00},\, \boldsymbol{\beta},\, \boldsymbol{\psi},\, \pi\big)
	\end{align*}
	and conditioning on the observed data $\mathbf{X}^{(1)}$ and 
	$\mathbf{X}^{(2)}$:
	\begin{align}
		\begin{split}
			&p\!\big(\mathbf{W}^{(1)}, \mathbf{W}^{(2)}, \mathbf{H},\,
			\boldsymbol{\alpha}^{(1)},\, p_{1,11}, p_{1,10}, p_{2,11}, p_{2,10},\,
			\gamma_{11}, \gamma_{00},\, \boldsymbol{\beta},\, \boldsymbol{\psi},\, \pi
			\mid \mathbf{X}^{(1)}, \mathbf{X}^{(2)}\big)
			\;\propto\; \\\
		&\underbrace{p\big(\mathbf{X}^{(1)}, \mathbf{X}^{(2)} \mid
				\mathbf{W}^{(1)}, \mathbf{W}^{(2)},
				\mathbf{H}\big)}_{\text{likelihood}}\times\;
			\underbrace{p\!\left(\mathbf{W}^{(1)} \mid
				\boldsymbol{\alpha}^{(1)}\right)
				p\!\left(\boldsymbol{\alpha}^{(1)}\right)}_{\mathbf{W}^{(1)}\text{ prior}}\\[4pt]
			&\times\; \underbrace{p\!\left(\mathbf{W}^{(2)} \mid \mathbf{W}^{(1)},
				\gamma_{11}, \gamma_{00}\right)
				p\!\left(\gamma_{11}\right)
				p\!\left(\gamma_{00}\right)}_{\mathbf{W}^{(2)}\text{ prior}}
			\times\; \underbrace{p\!\left(\mathbf{H} \mid \boldsymbol{\beta}\right)
				p\!\left(\boldsymbol{\beta} \mid \boldsymbol{\psi}\right)
				p\!\left(\boldsymbol{\psi} \mid \pi\right)
				p\!\left(\pi\right)}_{\mathbf{H}\text{ prior}}.
			\label{depe_posterior}
		\end{split}
	\end{align}
	The factorization in~\eqref{depe_posterior} reveals a conditional independence structure that yields closed-form full conditional distributions for each parameter block, derived in Appendix~\ref{app:conditionals}, enabling a Gibbs sampler in which the rows of the usage and pattern matrices can be updated independently within each sweep (Algorithm~\ref{alg:gibbs}).

	\subsection{Model Selection and Hyperparameter Tuning}
    For each candidate rank $R \in \{2, \ldots, R_{\max}\}$, we run four independent Gibbs chains from different random seeds and record the maximum log-posterior attained  by each. We then select  $R$ and the maximum a posteriori (MAP) iteration/chain that achieves the globally highest maximum log-posterior.  This criterion is a practical heuristic rather than a formal model-selection rule: the maximized joint density is not normalized over the parameter space, so its values are not strictly comparable between ranks and will tend to favor larger $R$ in the absence of a sparsity prior on $\mathbf{H}$. We therefore treat the selected rank as a working choice and return to this point in Section~\ref{summ_conclusions}. The priors on the noise parameters and loading matrices are uniform, letting the likelihood dominate, while the sparsity of $\mathbf{H}$  is encouraged through the spike-and-slab mixture on $\beta_g$ governed by $\psi_g$, with the overall sparsity level $\pi$ learned from the data. 
    
	\subsection{Computational Details}
	\label{subsec:computational_details}
    All models were fitted using a Gibbs sampler implemented in \texttt{R 4.5.2}. For each experiment and the real data analysis, we ran four chains from dispersed random initializations for $50,000$ iterations, discarding the first $10,000$ as burn-in, retaining all $40,000$ post-burn-in samples per chain ($160,000$ pooled samples). Convergence was assessed via log-posterior trace plots across chains. Code is available at 
     \url{https://github.com/wagala/jBBMF}.

	\subsection{Extension to \texorpdfstring{$M$}{M} Matrices}
	\label{subsec:M_matrices}
	
	Although Sections~\ref{subsec:prop_model} through~\ref{subsec:post_dist} are developed for two matrices ($i \in \{1, 2\}$), the \texttt{JBBMF} framework extends directly to $M \geq 2$ related binary matrices $\mathbf{X}^{(i)} \in \{0,1\}^{K \times G}$ for $i = 1, \ldots, M$ having a shared pattern matrix $\mathbf{H} \in \{0,1\}^{R \times G}$.  The Boolean product $Z^{(i)}_{kg} = \bigvee_{r=1}^{R}(W^{(i)}_{kr} \wedge H_{rg})$ remains as before while the full likelihood factorizes for all $M$ datasets as
	\begin{align*}
		p\!\left(\mathbf{X}^{(1)}, \ldots, \mathbf{X}^{(M)} \mid 
		\mathbf{W}^{(1)}, \ldots, \mathbf{W}^{(M)}, \mathbf{H}\right)
		= \prod_{i=1}^{M} \prod_{k=1}^{K} \prod_{g=1}^{G}\,
		p\!\left(X^{(i)}_{kg} \mid Z^{(i)}_{kg}\right),
	\end{align*}
	where each factor follows the Bernoulli noise model in~\eqref{Generative_Model2} with dataset-specific parameters $p_{i,11}$ and $p_{i,10}$. Each pair $(p_{i,11}, p_{i,10})$ receives an independent $\mathrm{Beta}(1,1)$ prior and its full conditional is a Beta distribution of the same form as 
	\eqref{distP1_11}, with counts computed from dataset $i$ alone.
	
    When the datasets are unordered, for example, $M$ independent cohorts 
	or experimental conditions, the loading matrices are treated as independent: for each $i$, $W^{(i)}_{kr} \mid \alpha^{(i)}_k 
		\sim \mathrm{Bernoulli}\!\left(\alpha^{(i)}_k\right),
		\qquad
		\alpha^{(i)}_k \sim \mathrm{Beta}(a_1, a_2) $, independently across $i = 1, \ldots, M$. For datasets are ordered in time, for example, $M$ sequential disease stages -- the conditional prior of Section~\ref{subsec:prior} extends to a first-order Markov structure: for $i = 2, \ldots, M$ and each $(k, r)$,
	\begin{align*}
		W^{(i)}_{kr} \mid W^{(i-1)}_{kr} \sim
		\begin{cases}
			\mathrm{Bernoulli}(\gamma_{11}), 
			& \text{if } W^{(i-1)}_{kr} = 1,\\[4pt]
			\mathrm{Bernoulli}(1 - \gamma_{00}), 
			& \text{if } W^{(i-1)}_{kr} = 0,
		\end{cases}
	\end{align*}
	with $W^{(1)}_{kr} \mid \alpha^{(1)}_k \sim 
	\mathrm{Bernoulli}(\alpha^{(1)}_k)$ and 
	$\alpha^{(1)}_k \sim \mathrm{Beta}(a_1, a_2)$ as before. The transition parameters $\gamma_{11}$ and $\gamma_{00}$ are shared between all consecutive pairs $(i-1, i)$, encoding a stationary transition mechanism; dataset-specific transition parameters could alternatively be introduced at the cost of additional hyperparameters. The prior on the shared pattern matrix $\mathbf{H}$ is unchanged 
	from Section~\ref{subsec:prior}. The Gibbs sampler extends without structural modification. The remainder of this paper focuses on the two-matrix case ($M = 2$) corresponding to the paired diagnosis-relapsed CNA application, for which Appendix~\ref{app:conditionals} derives the full conditional distributions.
	
	\section{Posterior Uncertainty Quantification}
	\label{sec:uncertainty}
	The Gibbs sampler produces posterior samples
	$\{(\mathbf W^{(i)(t)},\mathbf H^{(t)})\}_{t=1}^T$, with
	$\mathbf W^{(i)(t)}\in\{0,1\}^{K\times R}$ and
	$\mathbf H^{(t)}\in\{0,1\}^{R\times G}$, which we use to quantify uncertainty in the
	latent factor matrices and in the reconstructed Boolean matrices
	$\mathbf Z^{(i)(t)}=\mathbf W^{(i)(t)}\circ\mathbf H^{(t)}$. Throughout this section
	the superscript $(t)$ indexes the posterior draw and $(i)$ the dataset; the development
	is stated for a generic $i$ and applied to each dataset separately.

    This section briefly discusses factor-label switching, posterior inclusion probabilities for $\mathbf{H}$, uncertainty in the reconstructed matrix, and posterior summaries of the observation-noise parameters. Detailed derivations and further discussion are provided in Supplementary Section~S1.
	\subsection{Factor-label switching}
	
	The Boolean factorization is invariant to a common permutation of the columns of $\mathbf W$ and the corresponding rows of $\mathbf H$, analogous to the label-switching problem encountered in Bayesian latent-variable and mixture models \cite{stephens2000dealing,papastamoulis2016label}. We align sampled factors to a reference matrix $\mathbf H^{\mathrm{ref}}$ ($\mathbf H_{\text{truth}}$  in simulations; the MAP estimate in real data) using the Jaccard coefficient between factor patterns and the Hungarian algorithm \cite{kuhn1955hungarian} to find the optimal one-to-one assignment, then reorder rows of $\mathbf H^{(t)}$ and the corresponding columns of $\mathbf W^{(i)(t)}$ together.
		
    \subsection{Posterior inclusion probabilities for $\mathbf {H}$}
    Let each entry $H_{rg}\mid\mathbf X \sim \operatorname{Bernoulli}\!\left(\pi_{rg}^{H}\right)$, and so the posterior inclusion probability  is given as $ \pi_{rg}^{H} = \Pr(H_{rg}=1\mid\mathbf X)$ and is estimated as the proportion of aligned posterior samples with $H_{rg}^{*(t)}=1$.  The  posterior variance for  $H_{rg}$ is
		$\operatorname{Var}(H_{rg}\mid\mathbf X)=\pi_{rg}^{H}
		\left(1-\pi_{rg}^{H}\right),$ which is estimated by
		$\widehat{\operatorname{Var}}(H_{rg}\mid\mathbf X)=\widehat{\pi}_{rg}^{H}\left(1-\widehat{\pi}_{rg}^{H}\right).$ To enable easy intepretation  of $\widehat{\operatorname{Var}}(H_{rg}\mid\mathbf X)$, we define the normalized uncertainty score as  $\widehat U_{rg}^{H}=4\widehat{\pi}_{rg}^{H}\left(1-\widehat{\pi}_{rg}^{H}\right),$ which ranges from 0 to 1.
    
    \subsection{Uncertainty in the reconstructed matrix $\widehat{\mathbf Z}$}

    For each posterior draw we compute $\mathbf Z^{(i)(t)}=\mathbf W^{(i)(t)}\circ\mathbf H^{(t)}$ directly from unaligned samples (since alignment does not change the Boolean product), and estimate the posterior reconstruction probability $\pi_{kg}^{Z}$ and its normalized uncertainty $U_{kg}^{Z}$ analogously.
    
    \subsection{Posterior Summaries of the Observation-Noise Parameters}
    For each noise parameter $\mathbf{\theta}
    =\left\{p_{1,11},\,p_{1,10},\,p_{2,11},\,p_{2,10} \right\}$, we report the posterior mean and the equal-tailed $95\% $ credible interval $\mathrm{CrI}_{95\%}(\theta) =
    \left[ Q_{0.025}(\theta),\;Q_{0.975}(\theta)\right].$ computed from the pooled, burn-in-removed posterior samples.

\section{Simulation Experiments}
\label{sec:sim_Expts}
Three experiments are conducted, in which Experiments I and II reflect multi-condition settings, such as paired studies, in which the same samples are observed under different conditions. Here, the differences arise through changes in latent structure rather than in the observed dimensionality. Experiment III, on the other hand,  draws the latent factor matrices from a randomly chosen post-burn-in Gibbs iteration, producing block-sparse, overlapping activation patterns that approximate real CNA structure for multiple myeloma.

\subsection{Experimental Setups}
\label{subsec:exptl_set_ups}

In all experiments, each data set is represented as a binary matrix $\mathbf{X}^{(i)}_{\text{truth}} \in \{0,1\}^{K \times G}$, $i = 1, 2$,
where $K$ denotes the number of samples, $G$ the number of features, and we refer to each matrix $\mathbf{X}^{(i)}_{\text{truth}}$ as a condition. Each $\mathbf{X}^{(i)}_{\text{truth}}$ is generated as the Boolean product $\mathbf{X}^{(i)}_{\text{truth}} = \mathbf{W}^{(i)}_{\text{truth}} \circ \mathbf{H}_{\text{truth}}$, where $\mathbf{W}^{(i)}_{\text{truth}} \in \{0,1\}^{K \times R}$ is a sample-factor loading matrix and $\mathbf{H}_{\text{truth}} \in \{0,1\}^{R \times G}$ is a factor-feature pattern matrix. To simulate realistic noise, $10\%$ of the entries of each $\mathbf{X}^{(i)}_{\text{truth}}$ are selected at random and their values reversed, yielding the observed matrices $\mathbf{X}^{(1)}_{\text{Sim}}$ and $\mathbf{X}^{(2)}_{\text{Sim}}$.

The number of latent factors $R$ is fixed at $4$ in all experiments. The pattern matrix $\mathbf{H}_{\text{truth}}$ is held fixed in both conditions so that any differences between $\mathbf{X}^{(1)}_{\text{Sim}}$ and $\mathbf{X}^{(2)}_{\text{Sim}}$ arise solely through the dataset-specific loading matrices $\mathbf{W}^{(i)}_{\text{truth}}$. We construct rows of $\mathbf{H}_{\text{truth}}$ to exhibit block-structured patterns over features, inducing interpretable biclusters in the observed data. The size, density, and dispersion of these blocks are varied to reflect different regimes of latent structure. The loading matrices $\mathbf{W}^{(i)}_{\text{truth}}$ are varied across conditions to induce controlled patterns of overlap and condition-specific activation, thereby providing ground truth regarding which latent factors are shared across datasets. Heatmaps of the simulated data for Experiments~I--III are presented in Supplementary Figures~S1--S3, respectively.

Table~\ref{tab:sim_setup_summary} summarizes the design and purpose of each experiment. Full simulation details and ground-truth figures are provided in Supplementary Section~S2. 

\begin{table}[htbp]
\centering
\caption{Summary of the three simulation experiments. All experiments use
$R=4$ latent factors and a $10\%$ random entry-flip rate.}
\label{tab:sim_setup_summary}

\renewcommand{\arraystretch}{1.25}
\setlength{\tabcolsep}{6pt}

\begin{tabular}{p{1.4cm}p{2.3cm}p{6.0cm}p{3.3cm}}
\toprule
\textbf{Expt.} &
\textbf{Factor Structure} &
\textbf{Generative mechanism} &
\textbf{Purpose} \\
\midrule

I (S2.1) &
$2$ shared $+$ $2$ specific &
Dense block-structured $\mathbf{H}_{\mathrm{truth}}$.
$\mathbf{W}^{(2)}$ inherits $\mathbf{W}^{(1)}$ and gains
$12$ activations for each relapse-specific factor. &
Progression from diagnosis to relapse-specific latent factors. \\

\addlinespace

II (S2.2) &
$4$ factors &
$W^{(1)}_{kr}\sim\mathrm{Bernoulli}(0.3)$.
$\mathbf{W}^{(2)}$ is generated conditionally using
$p_{\mathrm{stay}}=0.8$ and $p_{\mathrm{gain}}=0.1$;
$\mathbf{H}_{\mathrm{truth}}$ is generated from
$\mathrm{Bernoulli}(0.2)$. &
Relapse conditional on $\mathbf{W}^{(1)}$ at diagnosis. \\

\addlinespace

III (S2.3) &
$4$ factors &
$\mathbf{W}^{(1)}$, $\mathbf{W}^{(2)}$, and $\mathbf{H}$
are taken from a single post-burn-in Gibbs iteration. &
Model-generated CNA-like structure. \\

\bottomrule
\end{tabular}

\vspace{2pt}
\begin{minipage}{\textwidth}
\footnotesize
\textit{Note:} All experiments use $K=62$ patients and
$G=44$ chromosomal arms.
\end{minipage}

\end{table}
\subsection{Evaluation Criteria and Diagnostics}
\label{subsec:diagnostics}
We assess performance using two complementary diagnostics:  (i) latent factor recovery by measuring how closely the estimated factor matrix $\hat{\mathbf{H}}$ reproduces the ground-truth factor matrix $\mathbf{H}_{\text{truth}}$ after Jaccard/Hungarian alignment of Section~\ref{sec:uncertainty}; and (ii)  reconstruction accuracy, evaluating how closely  $\hat{\mathbf{Z}}^{(i)} = \hat{\mathbf{W}}^{(i)} \circ \hat{\mathbf{H}}$ reproduces the 
ground-truth matrix $\mathbf{X}^{(i)}_{\text{truth}}$, measured by entry-wise metrics including F1 score, Matthews correlation coefficient (MCC), specificity, and reconstruction error rate. 

\section{Results for Simulation Experiments}
\label{sec:sim_results}

We apply \texttt{JBBMF} to each pair $(\mathbf{X}^{(1)}_{\text{Sim}},\mathbf{X}^{(2)}_{\text{Sim}})$ from Experiments~I--III to assess its ability to recover the ground-truth latent factors and reconstruct $\mathbf{X}^{(1)}_{\text{truth}}$ and $\mathbf{X}^{(2)}_{\text{truth}}$. Since no existing software implements a comparable joint Boolean matrix factorization, we use \texttt{Asso} \cite{Asso_4479462} applied independently to each matrix as a baseline. 

\subsection{Latent Variable Recovery}

For each estimated $\mathbf{H}$, we compute row-wise similarity against $\mathbf{H}_{\text{truth}}$ using the Jaccard index, then apply the Hungarian algorithm to find the optimal matching.

For Experiments~I--III, the $\mathbf{H}^{(1)}_{\texttt{Asso}}$ heatmaps (Supplementary Figures~S6(a)--S6(c)) show that \texttt{Asso} recovers at least one factor per dataset with high Jaccard similarity, but with noticeable cross-loading. In Experiment~I, one factor is recovered almost perfectly while others are partially mixed in all the estimated components. Experiment~II shows cleaner recovery but moderate off-diagonal values indicate residual ambiguity. In Experiment~III, diagonal similarities are uniformly high, yet non-negligible overlap suggests \texttt{Asso} splits or merges correlated structure. The $\mathbf{H}^{(2)}_{\texttt{Asso}}$ heatmaps (Supplementary Figures~S6(d)--S6(f)) reveal analogous behavior: diffuse similarity patterns in Experiment~I, clearer one-to-one matches in Experiment~II with some residual sharing, and strong diagonal dominance with moderate cross-factor overlap in Experiment~III. For $\mathbf{H}_{\texttt{JBBMF}}$ (Supplementary Figures~S6(g)--S6(i)), recovery improves progressively. In Experiment~I, one factor is recovered almost perfectly, while another loads on two components. In Experiment~II, the matrix is nearly a permutation matrix, with all diagonal Jaccard scores $\approx 1.0$ and minimal off-diagonal values. In Experiment~III, diagonal entries remain high  with moderate off-diagonal overlap.

\subsubsection{Similarity Between $\mathbf{H}^{(1)}_{\texttt{Asso}}$ and
$\mathbf{H}^{(2)}_{\texttt{Asso}}$}

\texttt{Asso} was applied independently to $\mathbf{X}^{(1)}_{\text{Sim}}$ and $\mathbf{X}^{(2)}_{\text{Sim}}$; consequently, each experiment yielded separate factor matrices,$\mathbf{H}^{(1)}_{\texttt{Asso}}$ and $\mathbf{H}^{(2)}_{\texttt{Asso}}$. To assess whether these independent factorizations recovered a common latent representation, we align their rows using the Hungarian algorithm and compute the Jaccard similarity
between corresponding factors. The independently estimated $\hat{\mathbf{H}}^{(1)}_{\texttt{Asso}}$ and $\hat{\mathbf{H}}^{(2)}_{\texttt{Asso}}$ show limited consistency across the three experiments. Although some matched factors exhibit high Jaccard similarity, many have only moderate or low overlap, indicating that
independent \texttt{Asso} factorizations may split or rearrange the latent structure in the two datasets. In contrast, \texttt{JBBMF} jointly estimates a single shared factor matrix, $\hat{\mathbf{H}}_{\texttt{JBBMF}}$, and therefore provides a common latent representation across both datasets. These results are presented Supplementary Figures~S6(j)--S6(l).

In addition to the factor estimates, we evaluated the recovery of the observation-model parameters under \texttt{JBBMF}. The posterior distributions of $(p_{1,11},p_{1,10},p_{2,11},p_{2,10})$ were concentrated near their true generating values across all three experiments, with small posterior standard deviations, indicating accurate recovery of the observation-model parameters by the Gibbs sampler. Detailed posterior summaries are presented in the Supplementary Table~S1.

\subsubsection{Reordered Estimated $\mathbf{H}$ Matrices}
\begin{figure}[tbp]
	\centering
	
	\subfloat[Experiment I.\label{fig:Hs_exp1}]{%
		\includegraphics[width=0.32\textwidth,height=5cm]{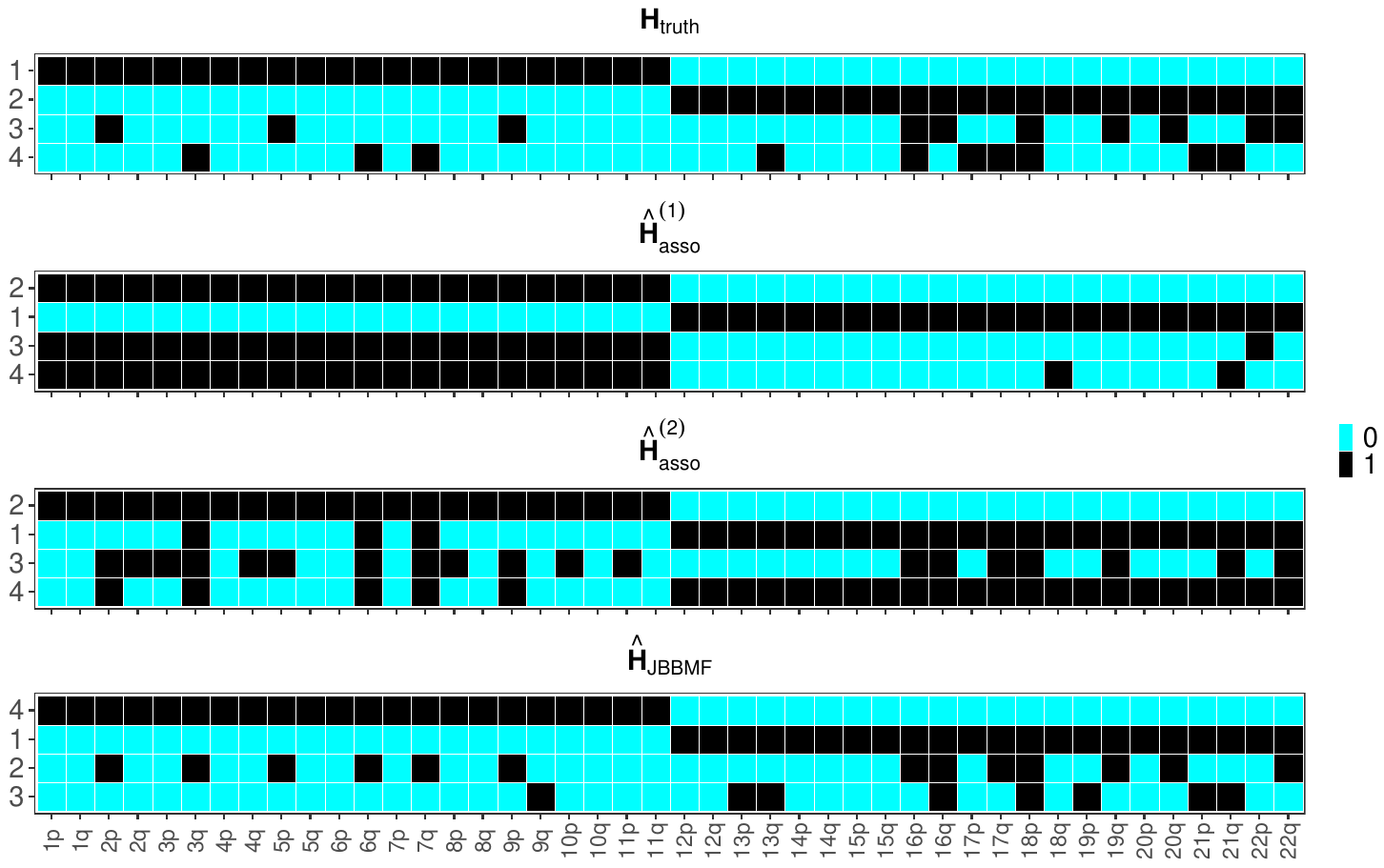}}
	\hfill
	\subfloat[Experiment II.\label{fig:Hs_exp2}]{%
		\includegraphics[width=0.32\textwidth,height=5cm]{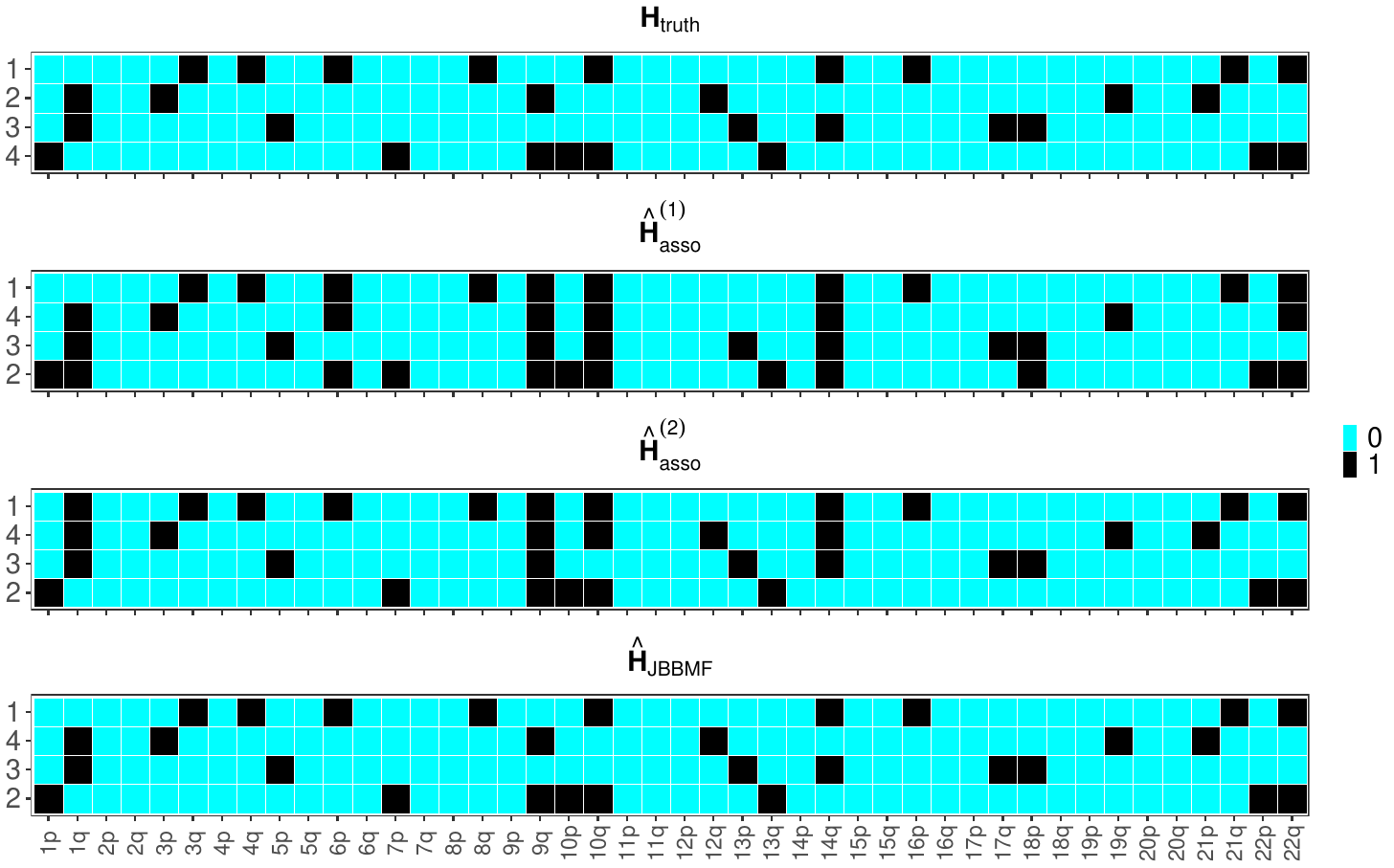}}
	\hfill
	\subfloat[Experiment III.\label{fig:Hs_exp3}]{%
		\includegraphics[width=0.32\textwidth,height=5cm]{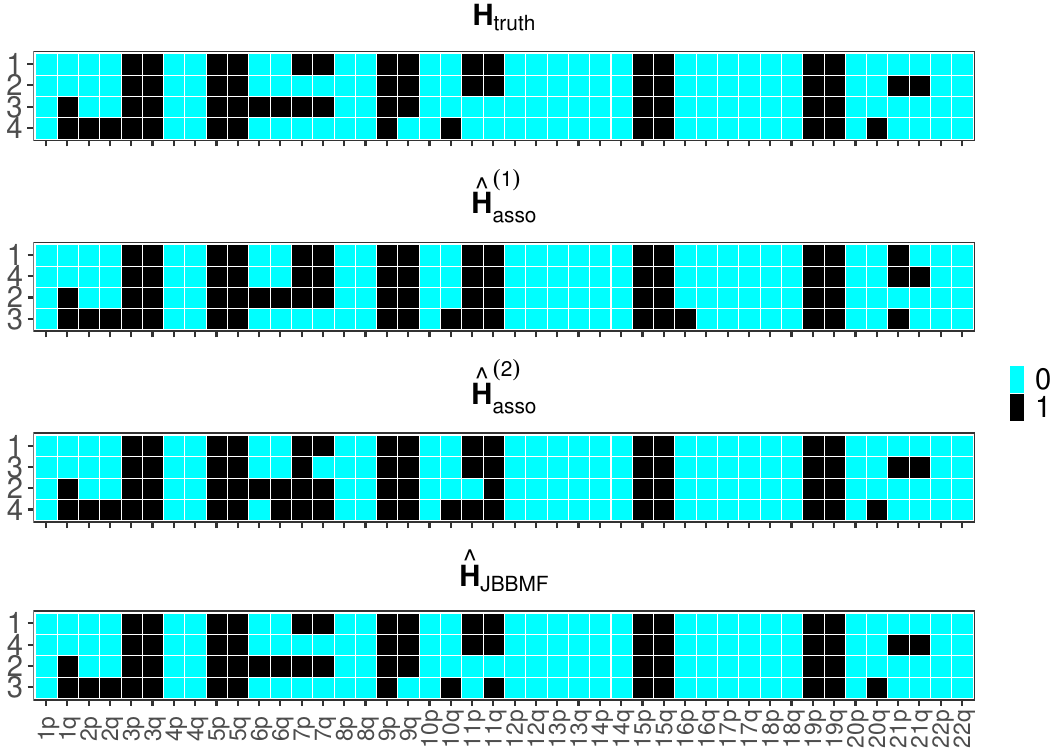}}
	
	\caption{Reordered $\hat{\mathbf{H}}$ matrices. In all the three experiments, $\hat{\mathbf{H}}_{\texttt{JBBMF}}$ recovers the true latent structure more accurately than the \texttt{Asso}-based estimates. \texttt{JBBMF} largely preserves the block structure, placement of nonzero entries, and factor-specific clusters in the three simulations, whereas the independent \texttt{Asso} estimates tend to misplace, split, or redistribute activity across factors.}
	\label{fig:all_Hs_expts}
	
\end{figure}

To visualize how closely the \texttt{Asso} and \texttt{JBBMF} factors reproduce the true latent patterns, we reorder the rows of each estimated matrices $\hat{\mathbf{H}}^{(i)}_{\texttt{Asso}}, i=1,2$ and $\hat{\mathbf{H}}_{\texttt{JBBMF}}$ to maximize alignment with the corresponding true factors and then display them as binary heatmaps. Figure~\ref{fig:all_Hs_expts} shows, for each experiment,four heatmaps arranged from top to bottom:  the true matrix $\mathbf{H}_{\text{truth}}$, the first estimate $\mathbf{H}^{(1)}_{\text{asso}}$, the second estimate $\mathbf{H}^{(2)}_{\text{asso}}$, and finally at the bottom the estimate $\mathbf{H}_{\texttt{JBBMF}}$.

\subsection{Reconstruction of $\mathbf{X}^{(1)}$ and $\mathbf{X}^{(2)}$ Matrices}
Table~\ref{tab:performance-metrics} reports four entry-wise metrics; specificity, F1 score, Matthews correlation coefficient (MCC) and reconstruction error rate for \texttt{JBBMF} and \texttt{Asso} applied to both matrices in each of the three experiments. Each metric treats the reconstruction of a binary 
matrix as a binary classification task, where the ground-truth entry $X^{(i)}_{kg,\text{truth}}$ is the class label and the reconstructed entry $\hat{Z}^{(i)}_{kg}$ is the prediction.
\begin{table}[tb]
	\centering
	\caption{Entry-wise reconstruction performance for Experiments~I--III. Specificity measures how well zeros are preserved; F1 and MCC summarize overall classification  accuracy; reconstruction error rate is the proportion of mismatched entries relative to the ground-truth matrix.}
	\label{tab:performance-metrics}
	\begin{tabular}{clcccc}
		\toprule
		Experiment & Method & Specificity & F1    & MCC   & Recon.~Error~Rate \\
		\midrule
		\multirow{4}{*}{I}
		& $\mathbf{X}^{(1)}_{\texttt{JBBMF}}$ & 0.983 & 0.980 & 0.971 & 0.012 \\
		& $\mathbf{X}^{(1)}_{\texttt{Asso}}$  & 0.997 & 0.996 & 0.995 & 0.002 \\
		& $\mathbf{X}^{(2)}_{\texttt{JBBMF}}$ & 0.986 & 0.922 & 0.886 & 0.051 \\
		& $\mathbf{X}^{(2)}_{\texttt{Asso}}$  & 0.948 & 0.924 & 0.883 & 0.054 \\
		\midrule
		\multirow{4}{*}{II}
		& $\mathbf{X}^{(1)}_{\texttt{JBBMF}}$ & 0.973 & 0.933 & 0.918 & 0.026 \\
		& $\mathbf{X}^{(1)}_{\texttt{Asso}}$  & 0.953 & 0.828 & 0.787 & 0.067 \\
		& $\mathbf{X}^{(2)}_{\texttt{JBBMF}}$ & 0.982 & 0.956 & 0.944 & 0.019 \\
		& $\mathbf{X}^{(2)}_{\texttt{Asso}}$  & 0.972 & 0.921 & 0.901 & 0.033 \\
		\midrule
		\multirow{4}{*}{III}
		& $\mathbf{X}^{(1)}_{\texttt{JBBMF}}$ &0.996 & 0.989& 0.986& 0.005 \\
		& $\mathbf{X}^{(1)}_{\texttt{Asso}}$  & 0.968 &0.942 & 0.925&0.027 \\
		& $\mathbf{X}^{(2)}_{\texttt{JBBMF}}$ & 0.996 &0.984&0.980 &0.007 \\
		& $\mathbf{X}^{(2)}_{\texttt{Asso}}$  &0.981& 0.955&0.942 &0.021 \\
		\bottomrule
	\end{tabular}
\end{table}
Both methods preserve zero entries well, achieving high specificity across all experiments. However, their relative performance depends on the underlying data structure.

In Experiment~I, \texttt{Asso}, Asso performs better on the first matrix, where the structure is simple and analyzed independently, whereas \texttt{JBBMF} performs better on the relapse matrix ($\mathbf{X}^{(2)}$) on three of the four metrics (specificity, MCC, and reconstruction error rate), with F1 scores that are effectively tied ($0.922$ for \texttt{JBBMF} vs. $0.924$ for \texttt{Asso}).

Generally, \texttt{Asso} is more competitive when matrices are simple and analyzed separately because it directly minimizes reconstruction loss. However, its performance declines when the matrices are interrelated. In contrast, \texttt{JBBMF} is more robust because it jointly models both matrices and exploits their shared structure, resulting in more accurate and consistent reconstructions.

\subsection{Distribution of the JBBMF reconstruction errors}
 Distributions of the reconstruction error counts across independent
		MCMC chains, fitted jointly to
		$\mathbf{X}^{(1)}_{\text{Sim}}$ and
		$\mathbf{X}^{(2)}_{\text{Sim}}$,
		for the latent variables
		$\mathbf{Z}^{(1)}$ and
		$\mathbf{Z}^{(2)}$
		in the three simulation experiments are presented in the Supplementary Figure~S7. 

 In Experiment I, error counts are notably higher for $\mathbf{Z}^{(2)}$ than $\mathbf{Z}^{(1)}$, with greater variability across chains. Experiment II shows elevated overall error counts with relatively consistent distributions between both latent variables. Experiment III produces the lowest error counts in general, with $\mathbf{Z}^{(2)}$ showing slightly reduced errors and tighter interquartile ranges compared to $\mathbf{Z}^{(1)}$. In all experiments, chains exhibit broadly comparable medians within each latent variable group, though outliers are present throughout, suggesting occasional reconstruction instability
\subsection{Posterior Inclusion Probabilities and Uncertainty for the Simulation Experiments}

\begin{figure}[tbp]
	\centering
	
	\subfloat[Posterior inclusion probabilities.\label{fig:H_probability}]{%
		\includegraphics[width=0.49\textwidth]{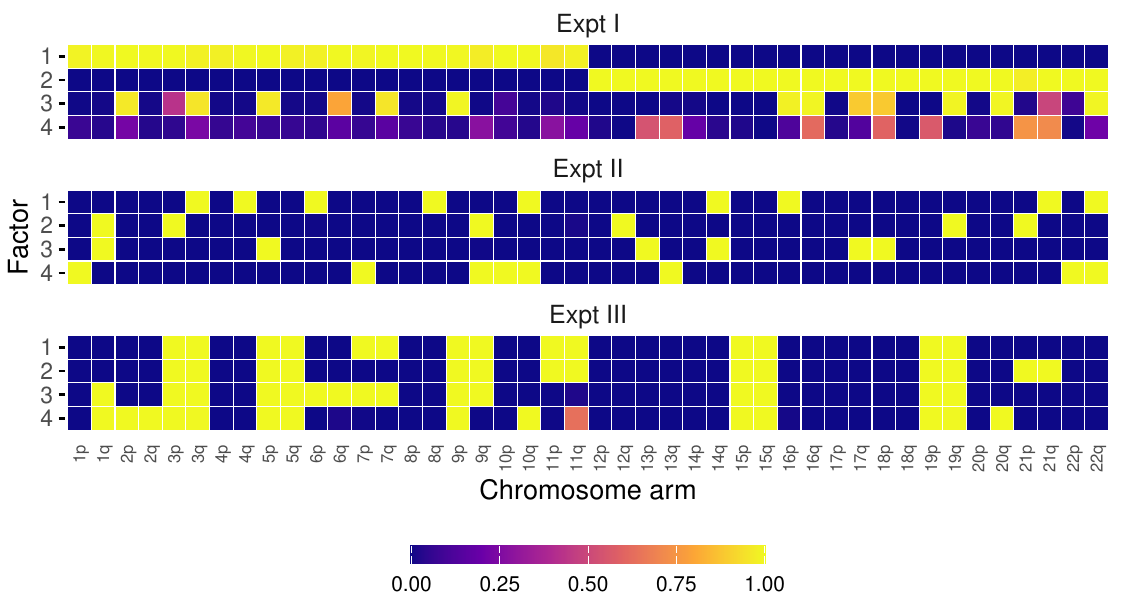}}
	\hfill
	\subfloat[Scaled posterior uncertainty.\label{fig:H_uncertainty}]{%
		\includegraphics[width=0.49\textwidth]{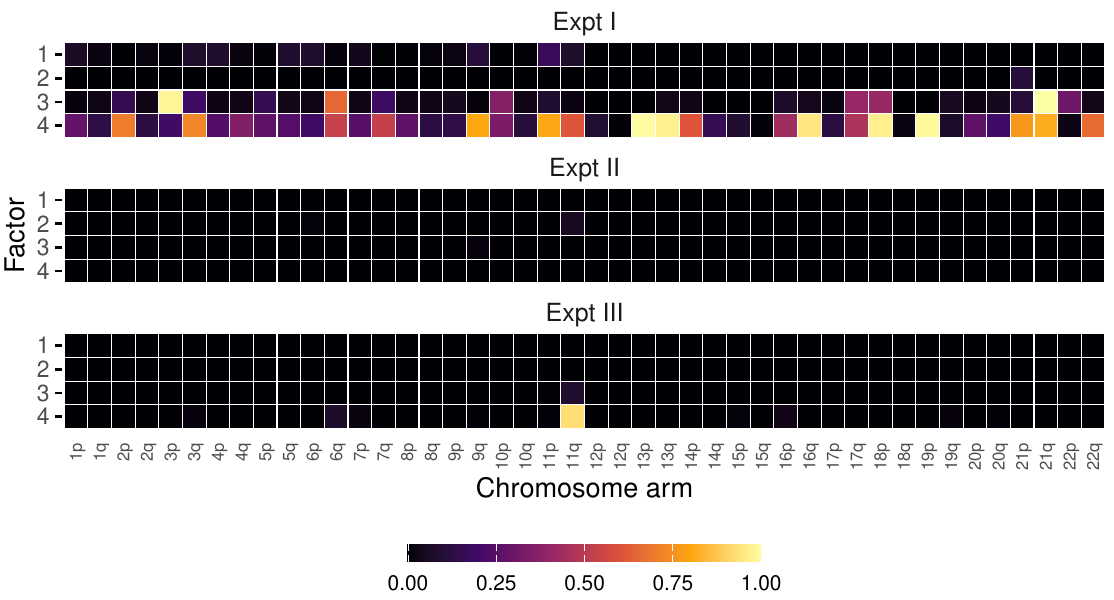}}
	
	\caption{
		Posterior summaries for the inferred latent factor matrix $\mathbf{H}$.
		(\textbf{a}) Posterior inclusion probabilities, where each cell represents
		the posterior probability that a chromosomal arm belongs to a latent factor.
		(\textbf{b}) Corresponding scaled posterior uncertainty,
		$U^{H}_{rg}=4\hat{\pi}^{H}_{rg}(1-\hat{\pi}^{H}_{rg})$,
		where values near $0$ indicate high posterior certainty and values near $1$
		indicate maximum uncertainty.}
	
	\label{fig:HPosteriorSummary}
	
\end{figure}

The posterior samples are first aligned to $\mathbf{H}_{\mathrm{truth}}$ following the alignment procedure described in Section~\ref{sec:uncertainty} before, these summaries are computed. Figure~\ref{fig:HPosteriorSummary} summarizes the posterior inclusion probabilities and corresponding uncertainty for the latent factor matrix $\mathbf{H}$ for the three simulation experiments. The inclusion probabilities are predominantly close to either $0$ or $1$, indicating strong posterior support for the inclusion or exclusion of chromosome arms within each latent factor. Consistent with these patterns, posterior uncertainty is generally low and is concentrated among the few assignments with intermediate inclusion probabilities. Experiment~II exhibits nearly uniform certainty, whereas Experiments~I and III show localized regions of elevated uncertainty.

\subsection{Clustering and Biclique Decomposition for Experiment~III}

To further evaluate the latent structure recovered by \texttt{JBBMF} in
Experiment~III, we examined the clustering patterns of the reconstructed
matrices and their corresponding biclique decompositions. These analyses
provide complementary views of the inferred structure: the clustered heatmaps
show the overall organization of patients and chromosome arms, while the
bicliques identify factor-specific co-alteration patterns.

For the clustering analysis, cosine distances were computed separately for
patients and chromosome arms, followed by complete-linkage agglomerative
clustering. The ordering obtained from each ground-truth matrix was applied to
both the ground-truth and corresponding \texttt{JBBMF} reconstruction.
Figure~S12 shows that the reconstructed matrices largely
preserve the block structures present in the ground-truth data, indicating that
\texttt{JBBMF} recovers the major patient and chromosome-arm groupings despite
the noise introduced in the simulation.

The recovered block structure is further characterized through the biclique
decomposition in Figure~S13. Each biclique is defined
by the outer product of a column of $\hat{\mathbf{W}}^{(i)}$ and the
corresponding row of $\hat{\mathbf{H}}$, representing a subgroup of patients
sharing alterations in a particular set of chromosome arms. The resulting
heatmaps reveal distinct co-alteration blocks of varying size and density,
with some overlap indicating that patients or chromosome arms may participate
in multiple latent patterns.

Supplementary Figures~S12 and
S13 demonstrate that \texttt{JBBMF} recovers the
major block structure of the simulated data while decomposing it into
interpretable factor-specific patient--chromosome-arm patterns.

\section{Application of JBBMF to the CNA Data}
\label{jBBMF_real_data}

We now apply the proposed \texttt{JBBMF} model to the paired multiple myeloma CNA data described in Section~\ref{subsec:copy_number_alterations}: binary matrices $\mathbf{X}^{(1)}$ and $\mathbf{X}^{(2)}$ recording chromosomal arm-level amplification for $K = 62$ patients at diagnosis and relapse across $G = 44$ chromosomal arms.

\subsection{Exploratory Analysis}

\begin{figure}[tbp]
	\centering
	\includegraphics[width=8cm]
	{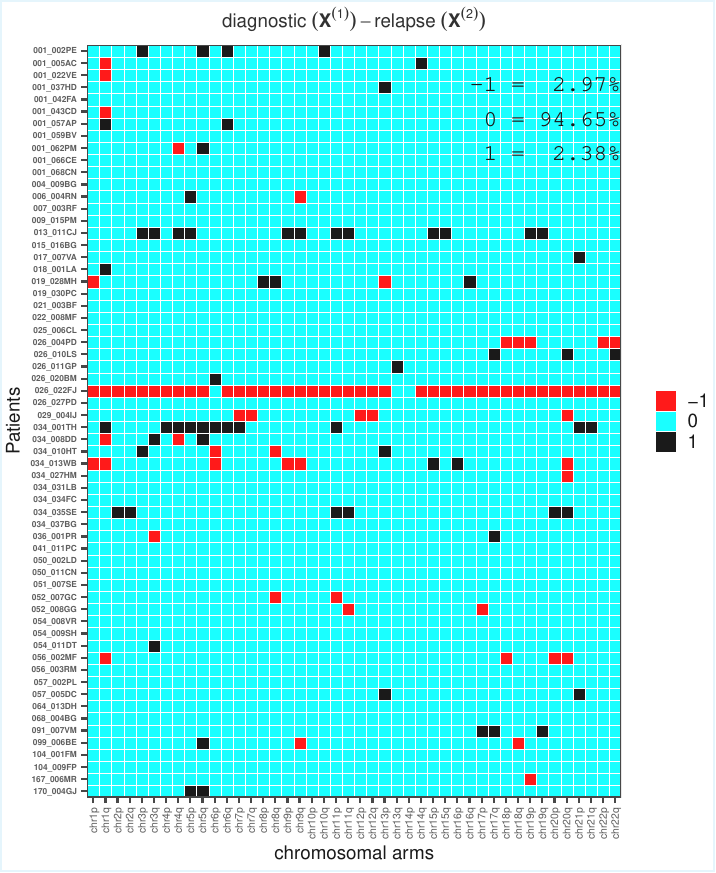}
	\caption{Differential CNA matrix 
		$\mathbf{X}^{(1)} - \mathbf{X}^{(2)}$ for the $K=62$ multiple myeloma patients across $G=44$ chromosomal arms. A value of $-1$ (red) indicates a loss of amplification from diagnosis to relapse; $0$ (cyan) indicates no change; $1$ (black) indicates a gain of 
		amplification at relapse.}
	\label{fig:CNA_MM_Exploratory}
\end{figure}

To explore the dynamics of CNA between the two stages of the disease before applying \texttt{JBBMF}, we complement the paired heat maps of Figure~\ref{fig:cna_diag_relapse} with the differential heat map of $\mathbf{X}^{(1)} - \mathbf{X}^{(2)}$ (Figure~\ref{fig:CNA_MM_Exploratory}) which highlights changes per-patient, per-arm between stages.

Most entries are zero, indicating relative genomic stability between the different time points for the majority of patients and chromosomal arms; scattered red and black tiles identify arms where amplification is lost or gained at relapse, reflecting clonal evolution of the tumor under selective pressure from therapy. This sparse but structured pattern of change motivates the joint factorization approach: a small number  of shared latent patterns should explain the bulk of the alteration 
profile at both stages, while stage-specific differences in how 
those patterns are used should account for the observed changes.

\subsection{Model Fitting and Selected Rank}

\begin{figure}[tbp]
	\centering
	
	\subfloat[Diagnosis.\label{fig:jbbmf_diag}]{%
		\resizebox{\textwidth}{!}{%
			\begin{tabular}{ccccc}
				\adjustbox{valign=m}{%
					\includegraphics[height=6cm,width=1.5cm]{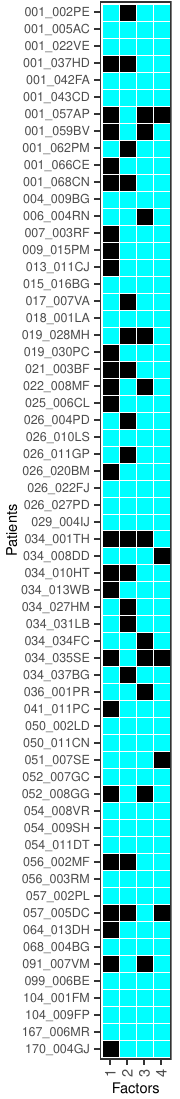}} &
				\adjustbox{valign=m}{$\circ$} &
				\adjustbox{valign=m}{%
					\includegraphics[height=.925cm,width=6cm]{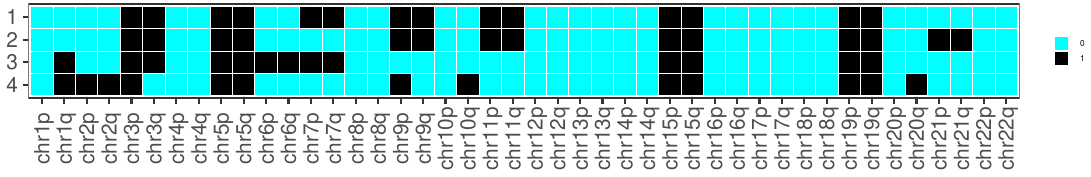}} &
				\adjustbox{valign=m}{$=$} &
				\adjustbox{valign=m}{%
					\includegraphics[height=6cm,width=6cm]{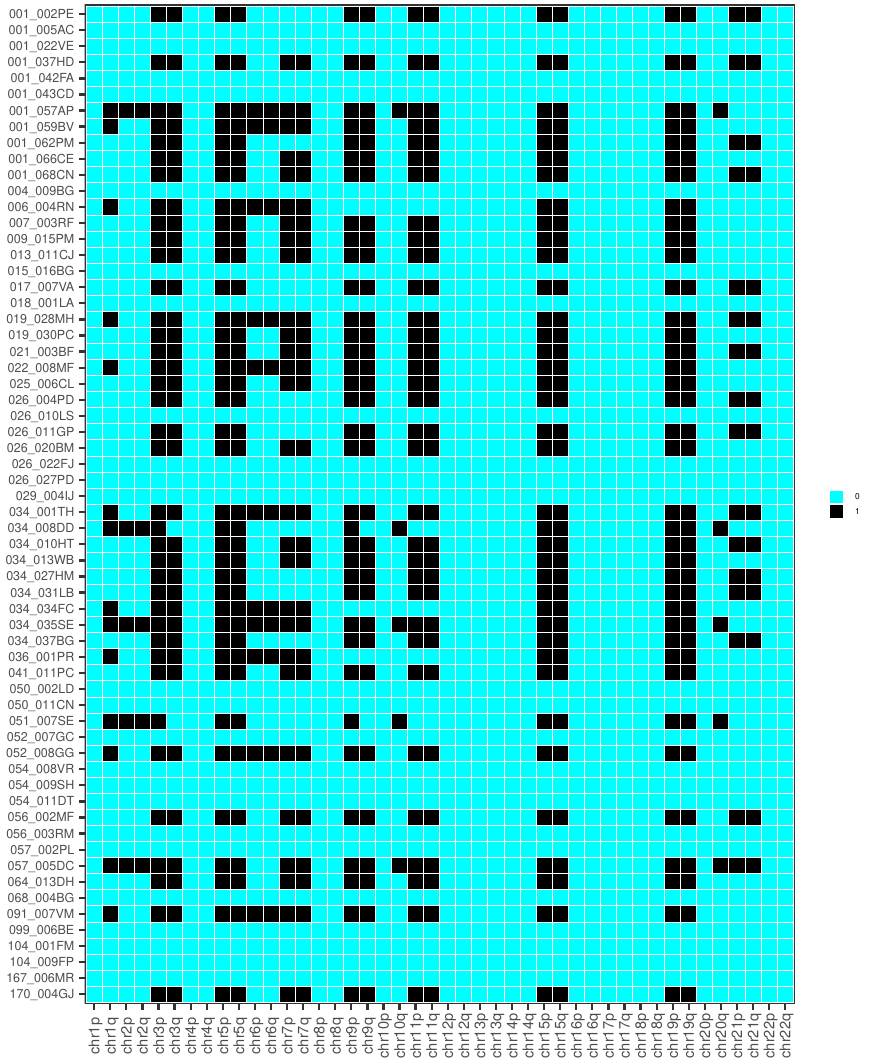}}\\[4pt]
				$\hat{\mathbf{W}}^{(1)}_{\text{diagnosis}}$ &&
				$\hat{\mathbf{H}}$ &&
				$\hat{\mathbf{Z}}^{(1)}_{\text{diagnosis}}$
	\end{tabular}}}
	
	\vspace{0.8cm}
	
	\subfloat[Relapse.\label{fig:jbbmf_relapse}]{%
		\resizebox{\textwidth}{!}{%
			\begin{tabular}{ccccc}
				\adjustbox{valign=m}{%
					\includegraphics[height=6cm,width=1.5cm]{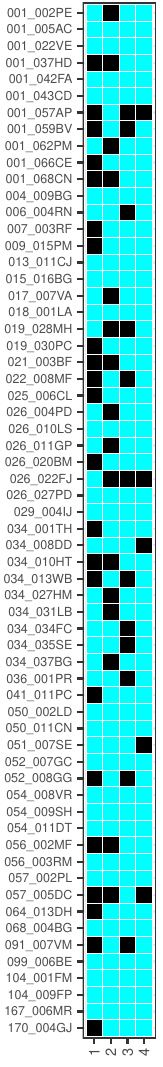}} &
				\adjustbox{valign=m}{$\circ$} &
				\adjustbox{valign=m}{%
					\includegraphics[height=.925cm,width=6cm]{H_CNA_MM.pdf}} &
				\adjustbox{valign=m}{$=$} &
				\adjustbox{valign=m}{%
					\includegraphics[height=6cm,width=6cm]{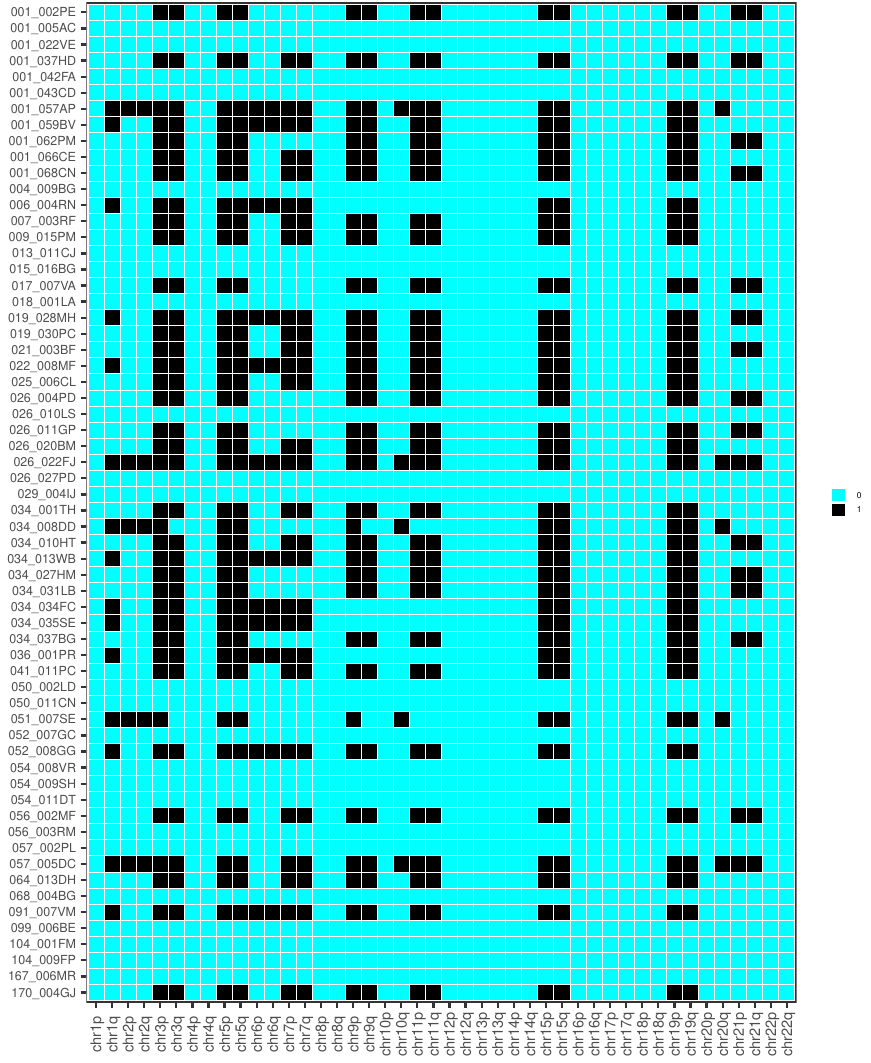}}\\[4pt]
				$\hat{\mathbf{W}}^{(2)}_{\text{relapse}}$ &&
				$\hat{\mathbf{H}}$ &&
				$\hat{\mathbf{Z}}^{(2)}_{\text{relapse}}$
	\end{tabular}}}
	
	\caption{\texttt{JBBMF} MAP factor matrices and reconstructions for the
		diagnosis--relapse CNA data ($R=4$). The top panel shows the diagnosis loading
		matrix $\hat{\mathbf{W}}^{(1)}$, the shared pattern matrix
		$\hat{\mathbf{H}}$, and the resulting reconstruction
		$\hat{\mathbf{Z}}^{(1)}$. The bottom panel shows the corresponding relapse
		loading matrix $\hat{\mathbf{W}}^{(2)}$ and reconstruction
		$\hat{\mathbf{Z}}^{(2)}$. The shared pattern matrix is identical in both
		panels; consequently, differences between the two reconstructions arise solely
		from the stage-specific loading matrices.}
	
	\label{fig:jbbmf_diagnosis_relapse}
	
\end{figure}

We applied \texttt{JBBMF} to $(\mathbf{X}^{(1)},\mathbf{X}^{(2)})$ for candidate ranks $R \in \{2,3,4,5,6\}$, running the Gibbs sampler and recording the maximum log-posterior attained at each rank. Of the ranks considered, $R=4$ yielded the highest maximum log-posterior and was therefore used in all subsequent analyses. The factor matrices from the iteration at which this maximum was attained were retained as the maximum a posteriori (MAP) estimates, and the reconstructed matrices, $\widehat{\mathbf{Z}}^{(1)}$ and $\widehat{\mathbf{Z}}^{(2)}$, were obtained through Boolean matrix multiplication (Figure~\ref{fig:jbbmf_diagnosis_relapse}). The reconstructions retain the dominant features of the observed data, including vertical bands of recurrently altered chromosome arms, differences in alteration burden across patients, and block-like subgroup structure, while reducing entry-level noise. Differences between the diagnosis and relapse reconstructions are confined to a small number of entries, consistent with the differential heatmap shown in Figure~\ref{fig:CNA_MM_Exploratory}.

\newpage

\subsection{Posterior Inclusion Probabilities and Uncertainty for the Multiple Myeloma Data}

\begin{figure}[tbp]
	\centering
	\includegraphics[width=0.85\textwidth]{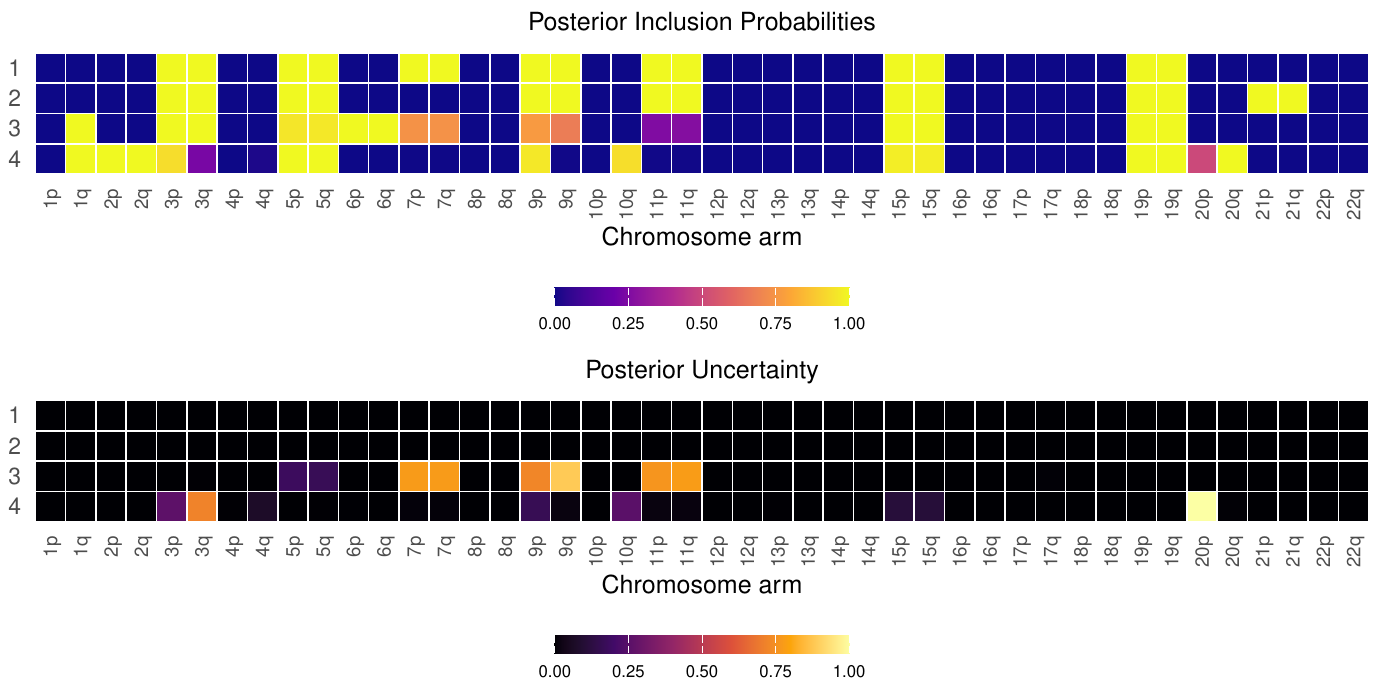} 
	\caption{Posterior inclusion probabilities (top) and corresponding scaled posterior uncertainties (bottom) for the inferred latent factor matrix H from the real multiple myeloma CNA data.}
	\label{fig:Hs_Prob_Inclusion_Real}
\end{figure}

Since true latent factors are unknown for real data, we quantify uncertainty from the posterior distribution obtained from the Gibbs sampler: posterior inclusion probabilities  and posterior reconstruction probabilities for for $\mathbf{H}$ (Figure ~\ref{fig:Hs_Prob_Inclusion_Real}) and $\hat{\mathbf{Z}}^{(i)}, i=1,2$ (supplementary Figure~ S11) respectively, each with a corresponding normalized uncertainty. Most chromosome arms exhibit inclusion probabilities close to $0$ or $1$, indicating well-identified, stable latent factors; only a small number of arms display moderate uncertainty. These summaries indicate that the inferred latent CNA signatures are highly stable and well supported by the data.

\subsection{Biological Interpretation, Clustering, and Biclique Decomposition}

The \texttt{JBBMF} decomposition provides three complementary levels of
interpretation. First, the shared CNA signatures, represented by the rows of
$\hat{\mathbf{H}}$ and estimated jointly from the diagnosis and relapse
matrices, encode recurring sets of co-altered chromosome arms. Because
$\hat{\mathbf{H}}$ is shared between the two disease stages, these signatures
represent common genomic configurations rather than stage-specific patterns
and help explain the persistent vertical bands in
Figure~\ref{fig:CNA_MM_Exploratory}.

Second, the stage-specific usage matrices
$\hat{\mathbf{W}}^{(1)}$ and $\hat{\mathbf{W}}^{(2)}$ describe how the shared
signatures are activated at diagnosis and relapse, respectively. A signature
with more patient activations in $\hat{\mathbf{W}}^{(1)}$ than in
$\hat{\mathbf{W}}^{(2)}$ is more prevalent at diagnosis, whereas the reverse
indicates greater prevalence at relapse. A signature that is nearly absent
from $\hat{\mathbf{W}}^{(1)}$ but active in
$\hat{\mathbf{W}}^{(2)}$ represents a shared CNA pattern whose patient-level
usage emerges or becomes more prominent at relapse.

\begin{figure}[tbp]
	\centering
	
	\subfloat[Diagnosis.\label{fig:X1_JBBooMF_CNA_Reordered}]{%
		\includegraphics[width=0.49\textwidth]{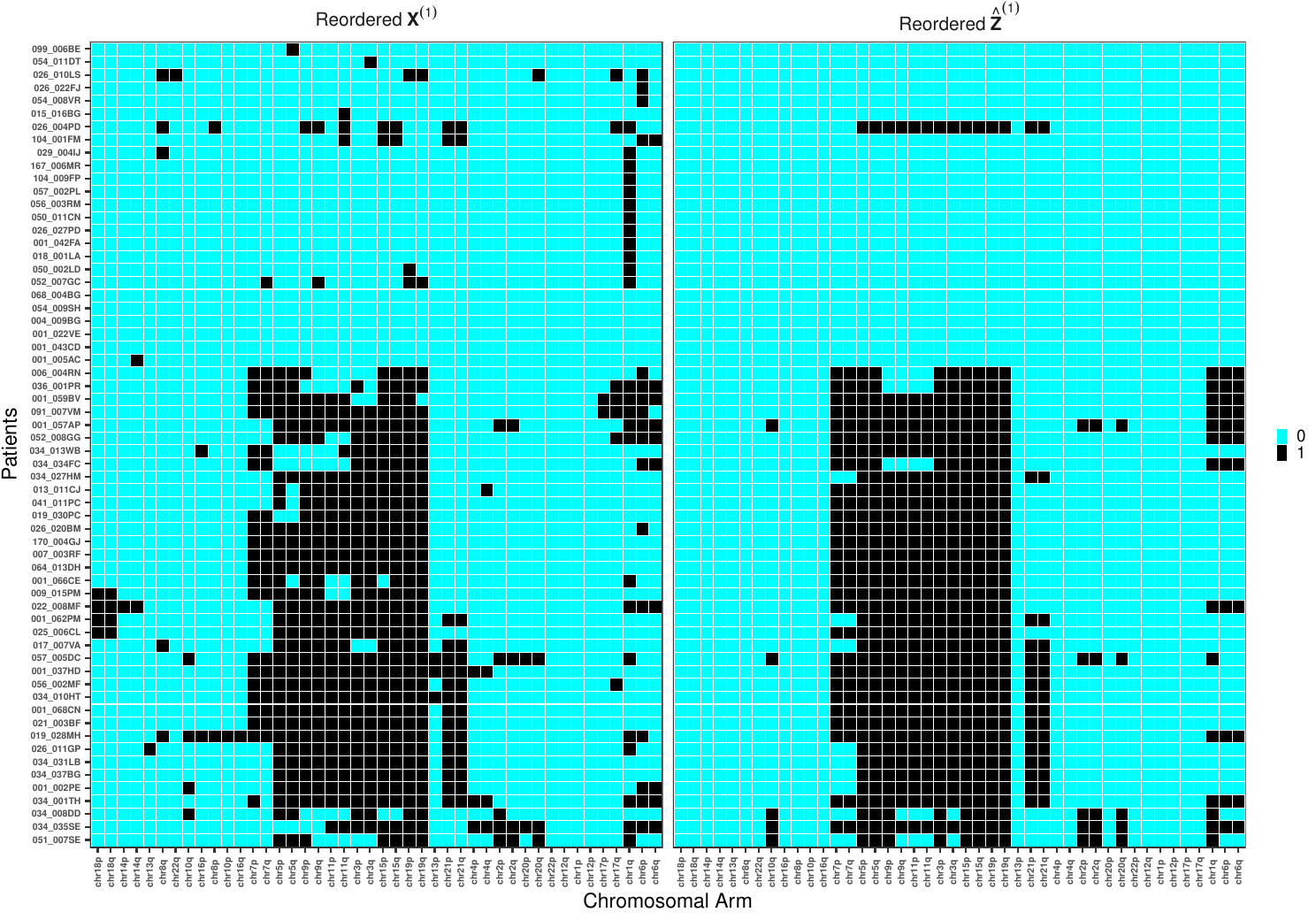}}
	\hspace{1mm}
	\subfloat[Relapse.\label{fig:X2_JBBooMF_CNA_Reordered}]{%
		\includegraphics[width=0.49\textwidth]{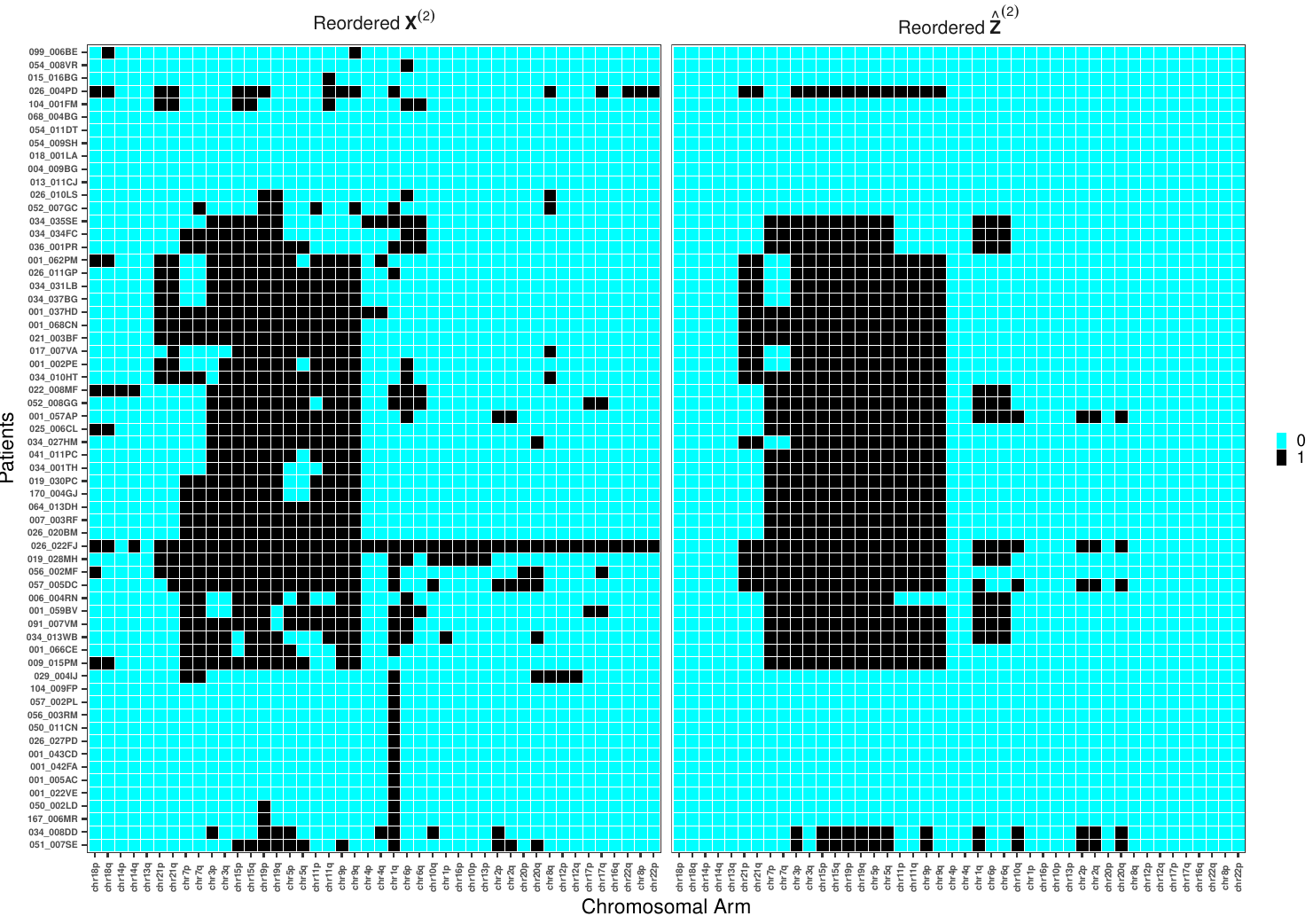}}
	
	\caption{Observed and \texttt{JBBMF}-reconstructed chromosomal alteration
		matrices at diagnosis (left) and relapse (right), reordered using a common
		clustering of patients and chromosome arms derived from the observed data.
		Within each panel, the left heatmap shows the observed binary CNA matrix, and
		the right heatmap shows the corresponding MAP reconstruction
		$\hat{\mathbf{Z}}^{(i)}$.}
	
	\label{fig:X1X2_JBBooMF_CNA_Reordered}
	
\end{figure}

Third, each row of $\hat{\mathbf{W}}^{(i)}$ identifies the signatures active
for a particular patient and therefore determines that patient's reconstructed
CNA profile. Patients with similar combinations of active signatures form
coherent subgroups. The resulting block structure is visible in
Figure~\ref{fig:CNA_MM_Exploratory}, supported by the common clustering of the
observed and reconstructed matrices in
Figure~\ref{fig:X1X2_JBBooMF_CNA_Reordered}, and made explicit by the biclique
decomposition (see Supplemenatary Figure~S14). Because the chromosome-arm composition of each signature is the same at both
stages, differences between the diagnosis and relapse bicliques arise from
changes in the patients activating each signature. These changes identify CNA
patterns whose prevalence remains stable from diagnosis to relapse, as well as those that become more or less prevalent at relapse.

\begin{figure}[tbp]
	\centering
	\includegraphics[width=0.65\textwidth]{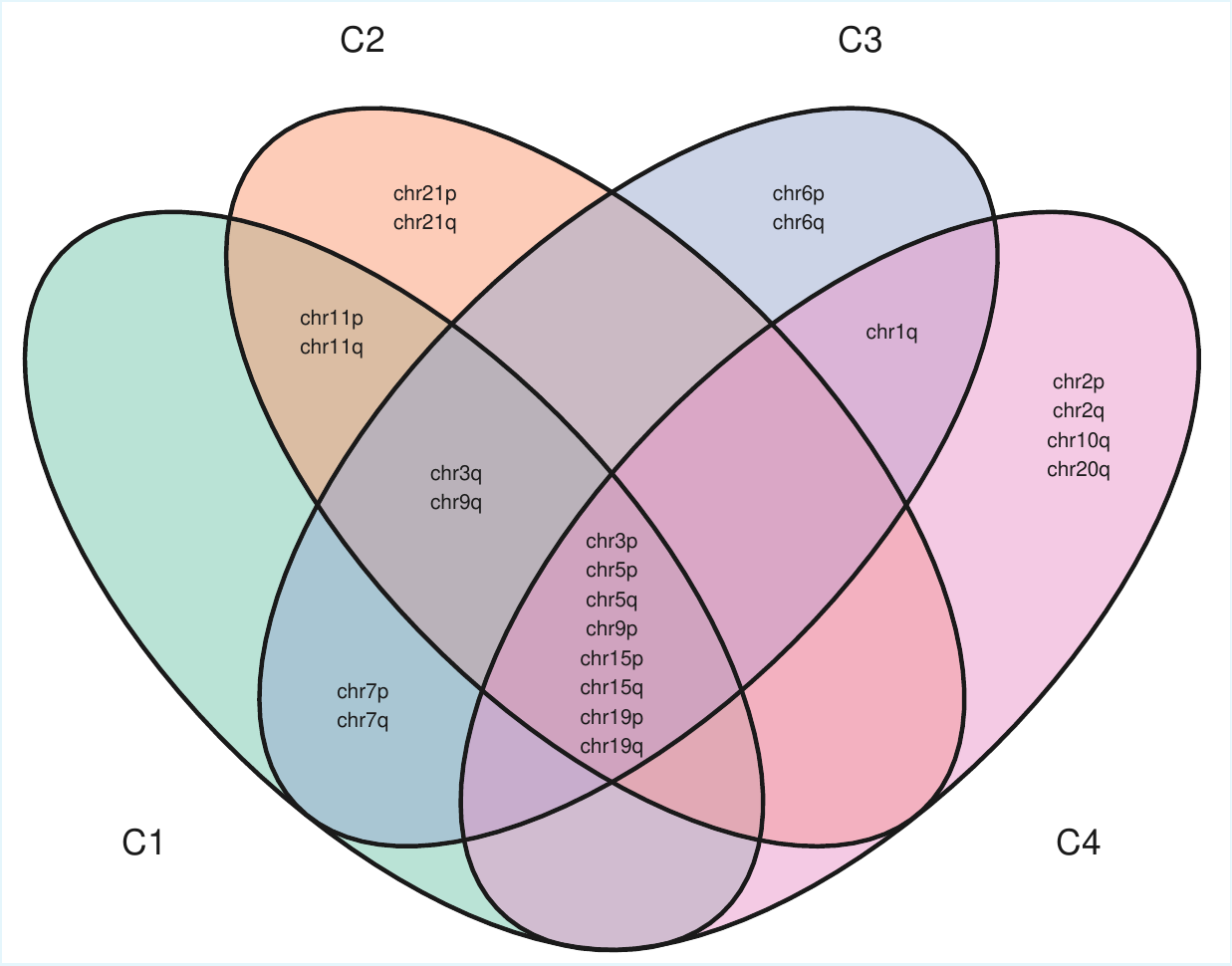}
	\caption{Venn diagram of chromosomal arms across the four 
		\texttt{JBBMF} latent factors. Each set C1, C2, C3, and C4 contains the arms for which $\hat{H}_{rg} = 1$ in factor $r$; overlapping regions identify arms shared between factors.}
	\label{fig:venn_cna_mm}
\end{figure}

Overlap among the four shared signatures is summarized by the Venn diagram in
Figure~\ref{fig:venn_cna_mm}. The chromosome arms
$\{\mathrm{chr3p},\mathrm{chr5p},\mathrm{chr5q},\mathrm{chr15p},\mathrm{chr15q},
\mathrm{chr19p},\mathrm{chr19q}\}$ occur in all four signatures, indicating that
they are common components of the inferred CNA architecture. In contrast,
arms unique to individual signatures distinguish the latent patterns and may
provide a basis for subsequent biological or clinical stratification of
patients.

\section{Summary and Conclusions}
\label{summ_conclusions}

We introduced a Joint Bayesian Boolean Matrix Factorization (\texttt{JBBMF}) model for analyzing multiple related binary matrices, with an emphasis on copy number alteration (CNA) data in multiple myeloma. The model simultaneously factorizes two binary matrices through a shared latent Boolean factor matrix and dataset-specific loading matrices, embedded within a fully probabilistic generative framework with a Bernoulli noise model and conjugate priors. This structure yields closed-form full conditional distributions and an efficient Gibbs sampler, supporting coherent uncertainty quantification and explicit noise modeling while preserving the interpretability of Boolean latent structure.

The simulation experiments compared \texttt{JBBMF} against \texttt{Asso} \citep{Asso_4479462} as the primary baseline. We note that the two joint Boolean matrix factorization methods reviewed in Section \ref{sec:Introduction}, namely \texttt{JSBMF} \citep{miettinen2012finding} and the relational approach of \cite{krmelova2013boolean} were not included as numerical baselines for the following reasons. \texttt{JSBMF} has no publicly available implementation, making direct 
numerical comparison infeasible. The relational approach of \cite{krmelova2013boolean} targets a different problem setting, linking independently estimated factorizations post-hoc rather than learning a shared latent representation, and does not provide a probabilistic noise model against which reconstruction metrics can be meaningfully compared. On the other hand,\texttt{Asso} 
is publicly available and represents the state-of-the-art deterministic approach to single-matrix Boolean factorization; comparing against it therefore provides a meaningful assessment of what is gained by moving to a joint 
probabilistic formulation. In all three simulation experiments, \texttt{JBBMF} recovered shared latent factors more accurately than independent \texttt{Asso} factorizations. It also achieved lower reconstruction error rates in every setting except the isolated, simply-structured diagnosis matrix of Experiment~I, where \texttt{Asso}---which directly minimizes Boolean reconstruction loss---was more accurate ($0.002$ versus $0.012$). The advantage of \texttt{JBBMF} is largest in the settings designed to encode conditional dependence between the two matrices.

Applied to CNA profiles at diagnosis and relapse in multiple myeloma, \texttt{JBBMF} decomposed the paired data into shared latent patterns and stage-specific usage matrices, providing an interpretable summary of genomic organization and its evolution over time. The Bayesian formulation naturally handles noise, supports the inclusion of prior biological knowledge, and yields posterior uncertainty over both the patterns and their usages. The biclique decomposition and cluster analysis 
identified a set of recurrently co-altered chromosomal arms common to both disease stages, revealed stage-specific changes in how those patterns are activated across patients, and highlighted coherent patient subgroups consistent with known 
biological heterogeneity in multiple myeloma.

The proposed \texttt{JBBMF} model has two main limitations. First, the number of latent factors $R$ is treated as fixed and selected by a MAP criterion rather than inferred from the data. A natural extension is to replace the finite-rank model with a nonparametric Bayesian formulation based on the Indian buffet process \citep{Griffiths2011}, which would allow $R$ to grow with the data and avoid the need for an explicit model selection step. Second, the model is developed for exactly two matrices; extending it to $M > 2$ matrices under 
a general prior hierarchy on the loading matrices, as outlined in Section~\ref{subsec:M_matrices}, remains to be fully implemented and evaluated. These extensions would broaden the applicability of the framework to longitudinal studies with more than two time points, multi-cohort genomic analyses, and other settings in which more than two related binary matrices 
are available.

 \section*{Acknowledgement(s)}
    This research was supported by the National Institutes of Health, National Cancer Institute (grant P01-155258; [G.P.\ and M.S.]), a gift from Anand Krishnamurthy \& Ruth Lievano ([G.P., M.S., and A.W.]), and the Paula and Rodger Riney Foundation ([G.P., M.S., and A.W.]). The content is solely the responsibility of the authors and does not necessarily represent the official views of the National Institutes of Health or the National Cancer Institute and does not necessarily reflect the views of Anand Krishnamurthy \& Ruth Lievano or the Paula and Rodger Riney Foundation.
\bibliographystyle{tfs}
\bibliography{references}
\appendix
 \clearpage
\addcontentsline{toc}{section}{Appendix}

\setcounter{section}{0}
\setcounter{figure}{0}
\setcounter{table}{0}
\setcounter{equation}{0}
\renewcommand{\thesection}{A\arabic{section}}
\renewcommand{\thesubsection}{A\arabic{section}.\arabic{subsection}}
\renewcommand{\thefigure}{A\arabic{figure}}
\renewcommand{\thetable}{A\arabic{table}}
\renewcommand{\theequation}{A\arabic{equation}}

\section{Conditional Distributions}
\label{app:conditionals}

Throughout this appendix, we write $Z^{(i)}_{kg} = \bigvee_{r=1}^R\!\left(W^{(i)}_{kr} 
\wedge H_{rg}\right)$ for the Boolean product at cell $(k,g)$ of dataset $i$, as defined in Section~\ref{sec:jBBMF}. All full conditionals are derived from the joint posterior in~\eqref{depe_posterior}. Since $\mathbf{W}^{(1)}$, $\mathbf{W}^{(2)}$, and $\mathbf{H}$ are binary 
matrices, their Gibbs updates proceed element-wise: for each element, the log-posterior is evaluated at $0$ and $1$, and a Bernoulli draw is made from the resulting probability.

\subsection{Full Conditional for $p_{1,11}$ and $p_{1,10}$}

Retaining only terms involving $p_{1,11}$, and noting that 
$Z^{(1)}_{kg} = 1$ defines the set of trials governed by $p_{1,11}$, 
the relevant part of the log-posterior is
\begin{align}
	\log p(p_{1,11} \mid \cdot) \propto 
	\sum_{k=1}^{K} \sum_{g=1}^{G} Z^{(1)}_{kg}
	\Big[ X^{(1)}_{kg} \log p_{1,11} 
	+ (1 - X^{(1)}_{kg}) \log(1 - p_{1,11}) \Big].
	\label{distp1_11a}
\end{align}
Let
\begin{align*}
	n^{(1)}_1 &= \sum_{k=1}^{K} \sum_{g=1}^{G} Z^{(1)}_{kg}\, X^{(1)}_{kg},
	\qquad
	n^{(1)} = \sum_{k=1}^{K} \sum_{g=1}^{G} Z^{(1)}_{kg}
\end{align*}
denote the number of successes and total number of trials under 
$p_{1,11}$, respectively. Under a uniform $\mathrm{Beta}(1,1)$ prior, 
equation~\eqref{distp1_11a} is proportional to a Beta kernel, giving
\begin{align}
	p_{1,11} \mid \cdot \;\sim\; 
	\mathrm{Beta}\!\left(n^{(1)}_1 + 1,\; n^{(1)} - n^{(1)}_1 + 1\right).
	\label{distP1_11}
\end{align}
Similarly, defining
\begin{align*}
	m^{(1)}_1 = \sum_{k=1}^{K}\sum_{g=1}^{G}(1-Z^{(1)}_{kg})\,X^{(1)}_{kg},
	\qquad
	m^{(1)} = \sum_{k=1}^{K}\sum_{g=1}^{G}(1-Z^{(1)}_{kg})
\end{align*}
as the successes and trials under $p_{1,10}$,
\begin{align}
	p_{1,10} \mid \cdot \;\sim\; 
	\mathrm{Beta}\!\left(m^{(1)}_1 + 1,\; m^{(1)} - m^{(1)}_1 + 1\right).
	\label{cond:p110}
\end{align}

\subsection{Full Conditional for $p_{2,11}$ and $p_{2,10}$}

By an identical argument applied to dataset 2, letting
\begin{align*}
	n^{(2)}_1 &= \sum_{k=1}^{K}\sum_{g=1}^{G} Z^{(2)}_{kg}\,X^{(2)}_{kg},
	\qquad
	n^{(2)} = \sum_{k=1}^{K}\sum_{g=1}^{G} Z^{(2)}_{kg},\\
	m^{(2)}_1 &= \sum_{k=1}^{K}\sum_{g=1}^{G}(1-Z^{(2)}_{kg})\,X^{(2)}_{kg},
	\qquad
	m^{(2)} = \sum_{k=1}^{K}\sum_{g=1}^{G}(1-Z^{(2)}_{kg}),
\end{align*}
we obtain
\begin{align}
	p_{2,11} \mid \cdot &\;\sim\; 
	\mathrm{Beta}\!\left(n^{(2)}_1 + 1,\; n^{(2)} - n^{(2)}_1 + 1\right),
	\label{distP2_11}\\
	p_{2,10} \mid \cdot &\;\sim\; 
	\mathrm{Beta}\!\left(m^{(2)}_1 + 1,\; m^{(2)} - m^{(2)}_1 + 1\right).
    \label{cond:p210}
\end{align}

\subsection{Full Conditional for $\alpha^{(1)}_k$}

Retaining terms involving $\alpha^{(1)}_k$ and letting 
$s_k = \sum_{r=1}^R W^{(1)}_{kr}$ denote the number of active factors 
for patient $k$ at diagnosis, the log full conditional is proportional 
to a Beta kernel, giving
\begin{align}
	\alpha^{(1)}_k \mid \cdot \;\sim\; 
	\mathrm{Beta}\!\left(s_k + a_1,\; R - s_k + a_2\right), 
	\qquad k = 1,\ldots,K.
	\label{cond:alpha}
\end{align}

\subsection{Full Conditionals for $\gamma_{11}$ and $\gamma_{00}$}

Using the counts $n_{11}$, $n_{10}$, $n_{01}$, $n_{00}$ defined in 
Section~\ref{subsec:prior}, and the Beta hyper-priors for $\gamma_{11}$ 
and $\gamma_{00}$, standard conjugacy gives
\begin{align}
	\gamma_{11} \mid \cdot &\;\sim\; 
	\mathrm{Beta}\!\left(n_{11} + u_{11},\; n_{10} + v_{11}\right),
    \label{cond:gamma11}\\
	\gamma_{00} \mid \cdot &\;\sim\; 
	\mathrm{Beta}\!\left(n_{00} + u_{00},\; n_{01} + v_{00}\right).
	\label{cond:gamma00}
\end{align}

\subsection{Log Full Conditional for $\bm{W}^{(1)}$}
\label{subsec:cond_logW1}
The log full conditional for $\bm{W}^{(1)}$ combines the likelihood 
contribution from dataset 1 with the  prior on $\bm{W}^{(1)}$ with the transition prior $p(\bm{W}^{(2)} \mid \bm{W}^{(1)})$:
\begin{align}
	\begin{split}
		\log p\!\left(\bm{W}^{(1)} \mid \cdot\right) \propto\;
		&\sum_{k=1}^{K}\sum_{g=1}^{G} \Big[
		Z^{(1)}_{kg}\,X^{(1)}_{kg}\log p_{1,11}
		+ Z^{(1)}_{kg}(1-X^{(1)}_{kg})\log(1-p_{1,11})\\
		&\quad + (1-Z^{(1)}_{kg})\,X^{(1)}_{kg}\log p_{1,10}
		+ (1-Z^{(1)}_{kg})(1-X^{(1)}_{kg})\log(1-p_{1,10})
		\Big]\\
		&+\sum_{k=1}^K\sum_{r=1}^R \Big[
		W^{(1)}_{kr}\log\alpha^{(1)}_k 
		+ (1-W^{(1)}_{kr})\log(1-\alpha^{(1)}_k)
		\Big]\\
		&+\sum_{k=1}^K\sum_{r=1}^R\Big[
		W^{(1)}_{kr}\big(W^{(2)}_{kr}\log\gamma_{11}
		+(1-W^{(2)}_{kr})\log(1-\gamma_{11})\big)\\
		&\quad+(1-W^{(1)}_{kr})\big(W^{(2)}_{kr}\log(1-\gamma_{00})
		+(1-W^{(2)}_{kr})\log\gamma_{00}\big)
		\Big].
	\end{split}
	\label{cond:W1}
\end{align}
Sampling proceeds element-wise for each $(k,r)$: the log-posterior is 
evaluated at $W^{(1)}_{kr} = 1$ and $W^{(1)}_{kr} = 0$, and 
$W^{(1)}_{kr}$ is drawn from the resulting Bernoulli distribution.

\subsection{Log Full Conditional for $\bm{W}^{(2)}$}
\label{subsec:cond_logW2}
The log full conditional for $\bm{W}^{(2)}$ combines the likelihood 
contribution from dataset 2 with the conditional prior 
$p(\bm{W}^{(2)} \mid \bm{W}^{(1)},\gamma_{11},\gamma_{00})$:
\begin{align}
	\begin{split}
		\log p\!\left(\bm{W}^{(2)} \mid \cdot\right) \propto\;
		&\sum_{k=1}^{K}\sum_{g=1}^{G} \Big[
		Z^{(2)}_{kg}\,X^{(2)}_{kg}\log p_{2,11}
		+ Z^{(2)}_{kg}(1-X^{(2)}_{kg})\log(1-p_{2,11})\\
		&\quad + (1-Z^{(2)}_{kg})\,X^{(2)}_{kg}\log p_{2,10}
		+ (1-Z^{(2)}_{kg})(1-X^{(2)}_{kg})\log(1-p_{2,10})
		\Big]\\
		&+\sum_{k=1}^K\sum_{r=1}^R\Big[
		W^{(1)}_{kr}\big(W^{(2)}_{kr}\log\gamma_{11}
		+(1-W^{(2)}_{kr})\log(1-\gamma_{11})\big)\\
		&\quad+(1-W^{(1)}_{kr})\big(W^{(2)}_{kr}\log(1-\gamma_{00})
		+(1-W^{(2)}_{kr})\log\gamma_{00}\big)
		\Big].
	\end{split}
	\label{cond:W2}
\end{align}
Sampling proceeds element-wise for each $(k,r)$.

\subsection{Log Full Conditional for $\bm{H}$}

The shared pattern matrix $\bm{H}$ enters through both datasets:
\begin{align}
	\begin{split}
		\log p\!\left(\bm{H} \mid \cdot\right) \propto\;
		&\sum_{k=1}^{K}\sum_{g=1}^{G} \Big[
		Z^{(1)}_{kg}\,X^{(1)}_{kg}\log p_{1,11}
		+ Z^{(1)}_{kg}(1-X^{(1)}_{kg})\log(1-p_{1,11})\\
		&\quad +(1-Z^{(1)}_{kg})\,X^{(1)}_{kg}\log p_{1,10}
		+(1-Z^{(1)}_{kg})(1-X^{(1)}_{kg})\log(1-p_{1,10})
		\Big]\\
		&+\sum_{k=1}^{K}\sum_{g=1}^{G}\Big[
		Z^{(2)}_{kg}\,X^{(2)}_{kg}\log p_{2,11}
		+Z^{(2)}_{kg}(1-X^{(2)}_{kg})\log(1-p_{2,11})\\
		&\quad +(1-Z^{(2)}_{kg})\,X^{(2)}_{kg}\log p_{2,10}
		+(1-Z^{(2)}_{kg})(1-X^{(2)}_{kg})\log(1-p_{2,10})
		\Big]\\
		&+\sum_{r=1}^R\sum_{g=1}^{G}\Big[
		H_{rg}\log\beta_g + (1-H_{rg})\log(1-\beta_g)
		\Big].
	\end{split}
	\label{cond:H}
\end{align}
Sampling proceeds element-wise for each $(r,g)$.
\subsubsection{Full Conditional for $\beta_g$}

For a fixed $g$, retaining only terms involving $\beta_g$ and 
conditioning on $\psi_g$, the log full conditional is
\begin{align}
\begin{split}
	&\log p(\beta_g \mid \cdot) \propto 
	\sum_{r=1}^R \Big[ H_{rg}\log\beta_g 
	+ (1-H_{rg})\log(1-\beta_g) \Big]\\
	&+ \log\!\left[
	\frac{\psi_g\,\beta_g^{b_1-1}(1-\beta_g)^{b_2-1}}{B(b_1,b_2)}
	+ \frac{(1-\psi_g)\,\beta_g^{c_1-1}(1-\beta_g)^{c_2-1}}{B(c_1,c_2)}
	\right].
    \end{split}
\end{align}
Letting $H_g = \sum_{r=1}^R H_{rg}$ and conditioning on $\psi_g$, 
the two cases simplify as follows.

\medskip
\noindent\textbf{Case $\psi_g = 1$:}
\begin{align*}
	\log p(\beta_g \mid \psi_g=1, \cdot) 
	&\propto H_g\log\beta_g + (R-H_g)\log(1-\beta_g)\\
	&+ (b_1-1)\log\beta_g + (b_2-1)\log(1-\beta_g) 
	- \log B(b_1,b_2)\\
	&= (H_g + b_1 - 1)\log\beta_g 
	+ (R - H_g + b_2 - 1)\log(1-\beta_g),
\end{align*}
since $-\log B(b_1,b_2)$ does not depend on $\beta_g$ and drops out.
This is the kernel of a $\mathrm{Beta}(H_g+b_1,\, R-H_g+b_2)$ distribution.

\medskip
\noindent\textbf{Case $\psi_g = 0$:}
\begin{align*}
	\log p(\beta_g \mid \psi_g=0, \cdot) 
	&\propto (H_g + c_1 - 1)\log\beta_g 
	+ (R - H_g + c_2 - 1)\log(1-\beta_g),
\end{align*}
which is the kernel of a $\mathrm{Beta}(H_g+c_1,\, R-H_g+c_2)$ distribution.

\medskip
Hence,
\begin{equation}
	\beta_g \mid \psi_g, \cdot \;\sim\;
	\begin{cases}
		\mathrm{Beta}\!\left(H_g + b_1,\; R - H_g + b_2\right), 
		& \text{if } \psi_g = 1,\\[6pt]
		\mathrm{Beta}\!\left(H_g + c_1,\; R - H_g + c_2\right), 
		& \text{if } \psi_g = 0.
	\end{cases}
	\label{cond:beta}
\end{equation}

\subsubsection{Full Conditional for $\psi_g$}

Unlike the conditional for $\beta_g$, updating $\psi_g$ requires 
comparing the two mixture components directly, so the Beta normalizing 
constants $B(b_1,b_2)$ and $B(c_1,c_2)$ do \emph{not} cancel and must 
be retained. The log full conditional is
\begin{align}
\begin{split}
	\log p(\psi_g \mid \beta_g, \pi) &\propto
	\log\!\left[
	\frac{\psi_g\,\beta_g^{b_1-1}(1-\beta_g)^{b_2-1}}{B(b_1,b_2)}
	+ \frac{(1-\psi_g)\,\beta_g^{c_1-1}(1-\beta_g)^{c_2-1}}{B(c_1,c_2)}
	\right]\\
	&+ \psi_g\log\pi + (1-\psi_g)\log(1-\pi).
    \end{split}
\end{align}
Evaluating at $\psi_g = 1$ and $\psi_g = 0$:
\begin{align*}
	l_1 &= (b_1-1)\log\beta_g + (b_2-1)\log(1-\beta_g) 
	- \log B(b_1,b_2) + \log\pi,\\
	l_0 &= (c_1-1)\log\beta_g + (c_2-1)\log(1-\beta_g) 
	- \log B(c_1,c_2) + \log(1-\pi).
\end{align*}
Note that omitting $-\log B(b_1,b_2)$ and $-\log B(c_1,c_2)$ 
from $l_1$ and $l_0$ respectively would yield an incorrect $q_g$ 
whenever $B(b_1,b_2) \neq B(c_1,c_2)$.
Hence $\psi_g \mid \text{rest} \sim \mathrm{Bernoulli}(q_g)$, where
\begin{align}
	q_g = \frac{\exp(l_1)}{\exp(l_1) + \exp(l_0)}.
	\label{cond:psi}
\end{align}

\subsubsection{Full Conditional for $\pi$}

Retaining only terms involving $\pi$ from the joint posterior in~\eqref{depe_posterior}, and using $\mathrm{Beta}(d_1,d_2)$ prior on $\pi$:

\begin{align}
	\log p(\pi \mid \text{rest}) 
	&\propto \sum_{g=1}^G \Big[\psi_g\log\pi 
	+ (1-\psi_g)\log(1-\pi)\Big]
	+ (d_1-1)\log\pi + (d_2-1)\log(1-\pi)\notag\\
	&= \left(\sum_{g=1}^G \psi_g + d_1 - 1\right)\log\pi
	+ \left(G - \sum_{g=1}^G \psi_g + d_2 - 1\right)\log(1-\pi),
\end{align}
which is the kernel of a Beta distribution. Hence,
\begin{align}
	\pi \mid \text{rest} \;\sim\; 
	\mathrm{Beta}\!\left(\sum_{g=1}^G \psi_g + d_1,\;
	G - \sum_{g=1}^G \psi_g + d_2\right).
	\label{cond:pi}
\end{align}

\subsection{The Gibbs Sampler for JBBMF}
\begin{algorithm}[H]
		\caption{Gibbs Sampler for \texttt{JBBMF}}
		\label{alg:gibbs}
		\begin{algorithmic}[1]
			
			\State \textbf{Initialize} all parameters.
			
			\For{$t = 1, \ldots, N$}
			
			\State Sample $p_{1,11},\, p_{1,10},\, p_{2,11},\, p_{2,10}$ 
			from Beta conditionals \eqref{distP1_11}--\eqref{distP2_11}.
			
			\For{$k = 1,\ldots,K$; $r = 1,\ldots,R$}
			\State Sample $W^{(1)}_{kr} \sim \mathrm{Bernoulli}
			\big(\sigma(l_1 - l_0)\big)$, \quad
			$l_v = \log p(W^{(1)}_{kr}=v \mid \text{rest})$,\; $v \in \{0,1\}$.
			\EndFor
			
			\For{$k = 1,\ldots,K$; $r = 1,\ldots,R$}
			\State Sample $W^{(2)}_{kr} \sim \mathrm{Bernoulli}
			\big(\sigma(l_1 - l_0)\big)$, \quad
			$l_v = \log p(W^{(2)}_{kr}=v \mid \text{rest})$,\; $v \in \{0,1\}$.
			\EndFor
			
			\For{$r = 1,\ldots,R$; $g = 1,\ldots,G$}
			\State Sample $H_{rg} \sim \mathrm{Bernoulli}
			\big(\sigma(l_1 - l_0)\big)$, \quad
			$l_v = \log p(H_{rg}=v \mid \text{rest})$,\; $v \in \{0,1\}$.
			\EndFor
			
			\For{$k = 1,\ldots,K$}
			\State Sample $\alpha^{(1)}_k \sim 
			\mathrm{Beta}\!\left(a_1 + s_k,\; a_2 + R - s_k\right)$,\quad
			$s_k = \sum_r W^{(1)}_{kr}$.
			\EndFor
			
			\State Sample $\gamma_{11} \sim \mathrm{Beta}(n_{11}+u_{11},\, n_{10}+v_{11})$,\quad
			$\gamma_{00} \sim \mathrm{Beta}(n_{00}+u_{00},\, n_{01}+v_{00})$.
			
			\For{$g = 1,\ldots,G$}
			\State $H_g \gets \sum_r H_{rg}$.\quad
			Sample $\beta_g \sim \mathrm{Beta}(b_1+H_g, b_2+R-H_g)$ if $\psi_g=1$,
			else $\mathrm{Beta}(c_1+H_g, c_2+R-H_g)$.
			\EndFor
			
			\For{$g = 1,\ldots,G$}
			\State $l_1 \gets \log\pi + (b_1-1)\log\beta_g + (b_2-1)\log(1-\beta_g)-\log B(b_1,b_2)$,
			\State $l_0 \gets \log(1-\pi) + (c_1-1)\log\beta_g + (c_2-1)\log(1-\beta_g)-\log B(c_1,c_2)$.
			\State Sample $\psi_g \sim \mathrm{Bernoulli}\big(\sigma(l_1-l_0)\big)$.
			\EndFor
			
			\State $n_\psi \gets \sum_g \psi_g$.\quad
			Sample $\pi \sim \mathrm{Beta}(d_1 + n_\psi,\, d_2 + G - n_\psi)$.
			
			\EndFor
			
			\State \Return posterior samples of all parameters.
			
		\end{algorithmic}
		\vspace{1mm}
		\footnotesize
		\emph{Note:} $\sigma(x)= \{ 1+\exp(-x) \}^{-1}$ denotes the logistic (sigmoid) function.
		
	\end{algorithm}
\end{document}